\documentclass[12pt,letterpaper]{article}

\usepackage[left=1in,right=1in,top=1in,bottom=1in]{geometry}
\usepackage{setspace}

\usepackage{amsmath,amssymb,amsthm,mathtools,bm}

\usepackage{graphicx}
\usepackage{adjustbox}
\usepackage{booktabs}
\usepackage{threeparttable}
\usepackage{longtable}
\usepackage{array}
\usepackage{multirow}
\usepackage{caption}
\usepackage{subcaption}
\usepackage{float}
\usepackage{placeins}
\usepackage{makecell}

\usepackage{enumitem}
\usepackage{xcolor}
\usepackage{appendix}
\usepackage{pdflscape}
\usepackage{csquotes}
\usepackage{bbm}

\usepackage{chngcntr}
\counterwithin{figure}{section}
\counterwithin{table}{section}

\usepackage[authoryear]{natbib}

\usepackage[colorlinks=true,linkcolor=blue,citecolor=blue,urlcolor=blue]{hyperref}
\usepackage[nameinlink,capitalize]{cleveref}

\theoremstyle{definition}

\title{Anatomy of the Market:\\A Body--Tail Test of Factor Models}

\author{Useong Shin\thanks{
		Sogang Business School, Sogang University (Seoul, Korea).\\
		ORCID: \href{https://orcid.org/0009-0003-0197-9003}{0009-0003-0197-9003}\\
		Email: \texttt{useong@sogang.ac.kr}
}}
\date{\today}

\begin{document}
	
	\maketitle
	\thispagestyle{empty}
	
	\begin{flushleft}
	\end{flushleft}
	
	
	\begin{abstract}
		
		In an ideal stochastic discount factor, zero pricing errors and maximum Sharpe ratio coincide; in a low-dimensional approximation they need not. I test this separation by decomposing an investible CRSP market into capitalization-ranked body and tail legs that recombine to the market return. At the daily frequency, all models pass the aggregate benchmark, but q5 alone leaves systematic offsetting leg alphas---negative body, positive tail---at all nine split ratios, despite holding the strongest spanning position. Matched random splits remove the pattern. Monthly aggregation attenuates q5's joint rejection and shifts relative weakness toward FF3, showing that internal consistency is frequency-dependent.
		
	\end{abstract}
	
	\pagenumbering{arabic}
	
	\newpage
	\section{Introduction}
	\label{sec:intro}
	
	In an ideal stochastic discount factor (SDF), two criteria coincide: the SDF prices every asset with zero alpha, and the tangency portfolio it spans attains the maximum Sharpe ratio. In a low-dimensional approximation they need not. \citet{BS17} make one side precise: zero-alpha tests depend on the chosen test assets while the maximum Sharpe ratio does not, so model comparison among traded-factor models should rest on spanning rather than on alphas over an arbitrary test-asset set. This paper takes the other side. A model can be strong on spanning and still leave systematic pricing errors on an economically ordered set of portfolios derived from the very market it prices in aggregate.
	
	The aggregate market regression is a natural starting point but a weak endpoint when the model already contains a market factor: a value-weighted market can carry a small alpha even when its components hold offsetting errors. I therefore decompose an investible CRSP market portfolio along its capitalization axis. The same market capitalization that sets each stock's weight in the aggregate also orders the decomposition---the body leg holds the dominant cumulative share of market value and the tail leg the remainder---and the two tradable, value-weighted legs dynamically recombine to the original market return.
	
	This makes the market portfolio its own testing device: the benchmark, universe, formation dates, and accounting stay fixed while only the internal test assets change. Because the recombination identity aggregates leg alphas back to the market alpha for every candidate model, a nonzero leg alpha is not by itself evidence against a model; the diagnostic content is the cross-model contrast in the leg-alpha pattern, which cannot be a mechanical consequence of splitting. The test asks whether pricing errors remain hidden along the capitalization axis once the market is compressed into a single aggregate return.
	
	I compare CAPM, FF3, Carhart, FF5, FF6, and q5, taking the daily frequency as the baseline---the native frequency of the CRSP panel and of the published daily factor series---and estimating body and tail alphas across nine cumulative-capitalization cutoffs, with matched random splits that hold the universe, dates, ratios, and accounting fixed but replace size-ranked assignment with random assignment. At the daily frequency every model prices the production market return $cMKT$ with an insignificant aggregate alpha, yet the body--tail tests separate them: CAPM, Carhart, FF5, and FF6 are broadly stable, while q5 rejects the joint zero-alpha restriction at all nine cutoffs with negative body and positive tail alphas. The contrast is sharp precisely because q5 is not weak on spanning---its ROE and expected-growth factors expand the mean--variance frontier beyond FF6---yet it leaves systematic offsetting alphas inside the market it prices. Matched random splits remove the pattern, locating it in the size-ranked decomposition rather than in the act of splitting.
	
	The evidence is frequency-dependent. Compounding the same daily leg returns to calendar months and refitting with monthly factors substantially attenuates the daily q5 joint rejection; in the common monthly sample the relative weakness shifts toward FF3, while q5 retains a weaker body-negative, tail-positive imbalance that surfaces mainly in alpha-sum tests. This is not a nuisance robustness result. Internal pricing consistency is conditional on the frequency at which portfolio and factor returns are evaluated, so factor models are not only model-dependent approximations of the SDF but frequency-dependent ones.
	
	A battery of daily diagnostics then locates the source. Alternative HAC lags, a $cMKT$ market-factor substitution, and external French and OSAP size deciles weigh against an inference-window artifact, a market-factor implementation issue, or a single test-asset construction. Factor-block ablations and a loading--premium decomposition point instead to q5's nonmarket profitability--growth block: the same block that contributes most to its spanning advantage generates a factor-implied body--tail gradient that runs against the realized tail-minus-body spread.
	
	The paper thus adds a third empirical object alongside aggregate-market fit and mean--variance spanning. The three are distinct: a model can pass the aggregate market regression, dominate on spanning, and still leave systematic internal pricing errors at a particular frequency. The cap-axis body--tail diagnostic makes that separation visible while holding the market identity fixed.
	
	\section{Theoretical Background and Related Literature}
	\label{sec:lit}
	
	\subsection{Two Criteria for Evaluating an Approximate SDF}
	\label{subsec:two-criteria}
	
	Linear factor models summarize common return variation with a few factors and price expected returns through exposures to them. Let $R^e_{i,t}$ be the excess return on test asset $i$ and $f_t$ a $K$-vector of factor returns; the standard time-series test estimates
	\begin{equation}
		R^e_{i,t}
		=
		\alpha_i
		+
		\beta_i'f_t
		+
		\varepsilon_{i,t},
		\qquad i=1,\ldots,N ,
	\end{equation}
	where the intercept $\alpha_i$ is the pricing error and the GRS test of \citet{GRS89} jointly tests $\alpha_i=0$ across the set. The same logic follows from the stochastic discount factor (SDF) approach, in which a valid SDF prices all traded assets and a factor model is a low-dimensional approximation to it \citep{Cochrane05,HJ97}. By \citet{KNS18}, the absence of near-arbitrage lets the SDF be represented through a few \emph{high-variance} return directions, so a low-dimensional model approximates the SDF along its dominant covariance directions and leaves pricing errors that are larger where expected-return variation is poorly aligned with those directions. Evaluation therefore concerns not only which anomalies a model explains but which pricing errors it leaves on a given asset set.
	
	For an \emph{ideal} SDF these distinctions vanish---the SDF that prices every asset is spanned by the maximum-Sharpe-ratio tangency portfolio---but for an approximation they come apart. \citet{BS17} make one side precise: for traded-factor models, comparing one against another reduces to whether each prices the other's factors, so common test-asset alphas add nothing and \emph{test assets are irrelevant for model comparison}, leaving factor spanning---linked to the maximum Sharpe ratio---as the criterion; \citet{BS18} develop the same mean--variance comparison. I take this as given. The body--tail exercise does not compare models by their alphas and is no substitute for spanning; it examines the other side of the gap---whether a model \emph{favored} on spanning still leaves systematic pricing errors on a particular, economically ordered set of test assets derived from the market it already prices.
	
	\subsection{Aggregate Market Alpha and Decomposition Diagnostics}
	
	Because every model here contains a market factor, a small alpha for the aggregate market portfolio is a sanity check rather than strong evidence about pricing elsewhere, especially when the market portfolio is measured from data close to the factor itself. It also does not mechanically imply small alphas on subsets of the same universe. In a fixed-weight case, let the market $M$ combine two tradable legs $B$ and $T$:
	\begin{equation}
		R^e_{M,t}
		=
		w_B R^e_{B,t}
		+
		w_T R^e_{T,t},
		\qquad
		w_B+w_T=1 .
	\end{equation}
	Under the same model the market alpha is the weighted average of the leg alphas,
	\begin{equation}
		\label{eq:alpha-agg}
		\alpha_M
		=
		w_B\alpha_B
		+
		w_T\alpha_T ,
	\end{equation}
	so $\alpha_M=0$ does not imply $\alpha_B=\alpha_T=0$: a nonzero leg alpha is not itself evidence against a model, and offsetting leg errors are consistent with a small aggregate alpha. Equation~\eqref{eq:alpha-agg} is an accounting identity that holds for \emph{every} model regardless of $f_t$, so any \emph{cross-model difference} in the leg pattern cannot be a mechanical consequence of the split; the test reads that asymmetry, not the presence of leg alphas, as informative. The weights vary over time in the buy-and-hold setting, but the same logic carries to the dynamic reconstruction below. This is why the exercise does not conflict with test-asset irrelevance, which concerns model \emph{comparison}: the body--tail design asks, one model at a time, whether it prices a size-ranked decomposition of the market it already prices in aggregate---internal pricing consistency, not mean--variance dominance.
	
	\subsection{Test-Asset Construction and the Position of This Paper}
	
	Empirical performance can depend on how test assets are constructed: when test assets and candidate factors share sorting rules, good fit can reflect both general pricing ability and alignment with the test-asset design. \citet{LM90} raise data-snooping concerns, \citet{LNS10} show inference is sensitive to test-portfolio structure, and \citet{GXZ25} stress the joint role of test assets and weak factors---the slack an approximate SDF leaves open, since by \citet{KNS18} different test assets weight different directions of misspecification. Body--tail legs are not anomaly portfolios sorted on book-to-market, profitability, investment, or momentum; they are size-ranked portfolios from the same investible universe whose value-weighted combination reconstructs the aggregate market return, fixing the benchmark they recombine to. A companion study \citep{Shin26} documents related construction dependence using CRSP-based random portfolios not presorted on characteristics; here the test assets are tied by construction to a market portfolio the models already price. I also use random splits, which hold everything fixed but assign stocks at random rather than by market-capitalization rank, as placebo benchmarks that separate the size-ranked rule from the act of forming two legs.
	
	I compare standard models for U.S. equity returns. The Fama--French three-factor model includes market, size, and value \citep{FF93}; the five-factor model adds profitability and investment \citep{FF15}; \citet{FF18} discuss factor selection; and the Carhart model adds momentum \citep{JT93,Carhart97}. The q-factor family is based on investment-based asset pricing: \citet{HXZ15} propose market, size, investment, and profitability factors, \citet{HMXZ19} weigh competing factors, and \citet{HMXZ21} add an expected-growth factor to form q5, with \citet{HXZ20} on anomaly replication and \citet{HMXZ24} on security-analysis information. On the spanning criterion these models are strong---as the tests below confirm, q5's profitability and expected-growth factors expand the mean--variance frontier well beyond the Fama--French factors, which makes q5 the informative case rather than a weak-model counterexample. The goal is not to name a single best model; a model can expand the opportunity set and still leave alphas on a specified set of portfolios. I therefore evaluate alphas on a body--tail decomposition rather than only on characteristic-sorted portfolios, benchmark them against matched random splits, and use cMKT substitutions and factor-block diagnostics to locate whether the pattern stems from market-factor implementation or a nonmarket block.
	
	\section{Data and Methodology}
	\label{sec:data}
	
	\subsection{CRSP Universe Selection}
	\label{subsec:crsp-universe}
	
	I build an investible U.S. equity universe from daily CRSP data over January 3, 1967 to December 31, 2024. The base sample is NYSE, AMEX, and NASDAQ common stocks. The screen is designed to retain almost all of the market's economic scale and trading activity while excluding the extreme microcap and illiquidity tails.
	
	Investibility is updated at each month-end using both market capitalization and liquidity. For stock $i$, month-end market capitalization is
	\begin{equation}
		ME_{i,t}
		=
		\left|PRC_{i,t}\right|
		\times SHROUT_{i,t}
		\times 1{,}000 .
	\end{equation}
	After sorting stocks in descending order of $ME$, the cumulative market-capitalization share of stock $i$ is
	\begin{equation}
		CumME_{i,t}
		=
		\frac{
			\sum_{j:ME_{j,t}\geq ME_{i,t}} ME_{j,t}
		}{
			\sum_j ME_{j,t}
		} .
	\end{equation}
	I measure liquidity by average dollar volume over the most recent 63 trading days ending at the month-end:
	\begin{align}
		DVOL_{i,d}
		&=
		\left|PRC_{i,d}\right|\times VOL_{i,d},
		\\
		ADV63_{i,t}
		&=
		\frac{1}{N_{i,t}}
		\sum_{d\in\mathcal{W}_t}DVOL_{i,d},
	\end{align}
	where $\mathcal{W}_t$ is the 63-day window ending at month-end $t$ and $N_{i,t}$ is the number of days in it with observed volume. I use monthly cross-sectional percentiles of $ADV63$ rather than an absolute cutoff, so the liquidity threshold adapts to a long sample with large changes in prices, volume, and market size.
	
	The selection rule uses hysteresis. An incumbent exits if $CumME_{i,t}>0.999$ or its $ADV63$ falls to the bottom 2.5\% of the monthly cross-section; an outside stock enters only if $CumME_{i,t}\leq0.995$ and its $ADV63$ reaches at least the fifth percentile, the stricter entry rule reducing boundary churn. Because early-sample NASDAQ volume history is limited, I defer the liquidity screen for a NASDAQ stock until 63 trading days of volume are available, while still applying the capitalization screen. All screening variables use only month-end information, and the universe set at a month-end applies from the next month's first trading day, preventing look-ahead bias.
	
	Across the full sample, the final universe averages 77.6\% of the common-stock base sample while preserving about 99.7\% of total market capitalization and 63-day dollar volume. At year-end 2024 it includes 2,426 of 3,804 base-sample stocks, with capitalization and dollar-volume preservation of 99.78\% and 99.28\%. The screen thus removes the extreme illiquidity tail while retaining nearly all of the market's economic weight.
	
	\subsection{Market-Factor Replication}
	\label{subsec:market-replication}
	
	The tests start from an investible market portfolio built directly from CRSP individual stocks, so I first check whether later alpha estimates could be driven by market-return implementation differences. I construct a daily value-weighted market return, $cMKT$, from the universe above; it is both the benchmark that the body--tail and random-split portfolios reconstruct and the series used to gauge implementation gaps against standard market factors.
	
	Replicating a broad market return is not merely value-weighting all stocks: delisting and dividend-inclusive returns, formation dates, coverage, and the treatment of small or illiquid stocks introduce small daily differences that accumulate over decades and affect intercepts. I compute $cMKT$ as a daily value-weighted buy-and-hold portfolio with the universe fixed at each month-end and applied from the next month's first trading day, weights based on formation-date market capitalization, and returns including dividends, delisting returns where observed, and proportional reinvestment of within-portfolio cash flows into surviving securities. The aim is not to reverse-engineer any provider's factor but to ask whether an independently constructed CRSP market return is close enough to the Fama--French and q5 market factors for the decomposition tests to be meaningful.
	
\begin{figure}[H]
		\centering
		\includegraphics[width=0.95\linewidth]{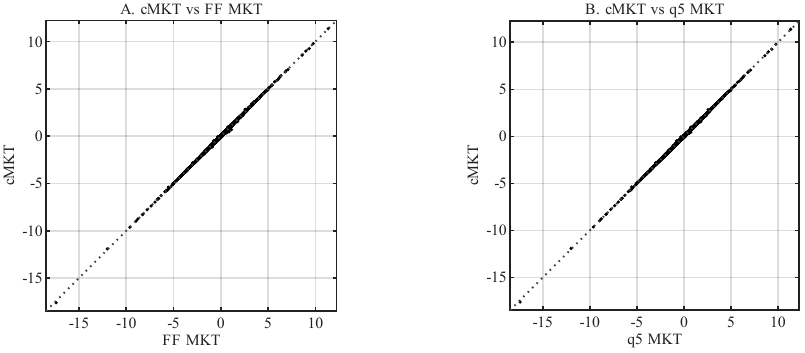}
		\caption{CRSP-based market return and standard market factors}
		\label{fig:mkt-replication}
		\begin{minipage}{0.95\linewidth}
			\footnotesize
			\emph{Note:} The figure reports daily scatter plots of $cMKT$, the market return that I construct from the CRSP investible universe, against the Fama--French and q5 market returns. The comparison uses total market returns after adding each provider's risk-free rate back to the excess market return. The dashed line is the 45-degree line. The common sample runs from January 3, 1967 to December 31, 2024.
		\end{minipage}
	\end{figure}

	It is. Over the common sample, $cMKT$ correlates with the Fama--French market return at 0.999778 ($R^2=0.999556$; daily RMSE 2.21 bp, mean absolute error 1.35 bp) and with the q5 market return at 0.999799 ($R^2=0.999598$; RMSE 2.10 bp, MAE 1.27 bp), with beta near one and an economically small intercept in both. The q5 factors used below are the providers' daily series \citep{globalqFactors}, not a frequency conversion of my own, so the daily frequency is not imposed on q5---it is one of the frequencies at which the provider releases the model. Figure~\ref{fig:mkt-replication} shows most observations near the 45-degree line.
	
	\FloatBarrier
	\FloatBarrier
	
	High accuracy does not make construction trivial; long-run alpha tests remain sensitive to delisting, reinvestment, formation dates, coverage, and illiquid-stock inclusion. I therefore treat the $cMKT$ comparison as a diagnostic that separates implementation-gap alphas from those in the test portfolios. As a sanity check, since each model contains a market factor, the intercept should be near zero when the full investible market is the test asset. Table~\ref{tab:market-sanity-check} regresses the full market return on each model: all have $R^2$ above 0.999 and every alpha $p$-value exceeds 0.05 under a 21-day Newey--West lag.
	
	\FloatBarrier
	\begin{table}[H]
		\centering
		\singlespacing
		\caption{Market sanity check for the full market return}
		\label{tab:market-sanity-check}
		\footnotesize
		\setlength{\tabcolsep}{4.8pt}
		\begin{tabular}{lrrrrr}
			\toprule
			Model & Observations & Annual alpha (bp) & $t(\alpha)$ & $p(\alpha)$ & $R^2$ \\
			\midrule
			CAPM    & 14,598 & -1.02 & -0.22 & 0.829 & 0.999556 \\
			FF3     & 14,598 &  0.21 &  0.05 & 0.963 & 0.999571 \\
			Carhart & 14,598 & -0.06 & -0.01 & 0.989 & 0.999571 \\
			FF5     & 14,598 & -1.07 & -0.24 & 0.812 & 0.999572 \\
			FF6     & 14,598 & -1.18 & -0.26 & 0.792 & 0.999572 \\
			q5      & 14,598 & -5.78 & -1.44 & 0.151 & 0.999619 \\
			\bottomrule
		\end{tabular}
		
		\vspace{0.4em}
		\parbox{0.95\textwidth}{\footnotesize
			\emph{Note:} The table reports regressions of the full investible market return constructed in this paper on each candidate factor model. Alpha is the daily intercept multiplied by 252 and reported in basis points per year. All tests use the common sample from January 3, 1967 to December 31, 2024. The $p$-values use a Newey--West lag of 21 trading days.}
	\end{table}
	\FloatBarrier
	
	The pass is a consistency gate, not a claim that implementation is irrelevant or that any portfolio from the same universe must price at zero. It establishes only that all models price the aggregate market closely under the baseline test, so later leg-level patterns read as properties of the body--tail and random-split portfolios rather than failures of the benchmark. Because $cMKT$ is empirically close to both provider market returns, simple market-factor implementation error is an unlikely standalone explanation for the body--tail results; as a more direct control, I also replace each provider's market factor with $cMKT$ in Section~\ref{sec:diagnostics}.
	
	\subsection{Observation Frequency and Monthly Aggregation}
	\label{subsec:frequency-design}
	
	The baseline analysis is daily because the CRSP return panel is observed daily; this uses the native frequency of the stock-level data and is also natural for the reconstruction, whose body and tail weights evolve through daily returns, delisting returns, and reinvested cash flows and whose identity is verified at the daily level. Daily estimation is not the only relevant implementation, however: many asset-pricing applications run monthly, and both factor providers distribute monthly series. I therefore treat daily as the primary design and add a monthly aggregation as a complementary frequency check, asking whether the same economically ordered decomposition produces the same internal pricing pattern at the lower frequency.
	
	For each leg I compound daily returns within each calendar month,
	\begin{equation}
		R_{\ell,p,m}
		=
		\prod_{d\in m}\left(1+R_{\ell,p,d}\right)-1,
		\qquad
		\ell\in\{B,T\} ,
	\end{equation}
	and aggregate the full market and random-split legs the same way. Monthly excess returns use the corresponding monthly risk-free rate, and regressions use monthly factor returns; where provider monthly series exist I use them rather than compounding provider daily factors, so the monthly test is a genuine lower-frequency implementation of each model rather than a mechanical aggregation of the daily regression. The monthly tests hold the universe, June formation schedule, split ratios, buy-and-hold accounting, and assignment rule fixed, so the only substantive change is the observation frequency. Inference uses Newey--West standard errors with 21 trading-day lags (daily) or six monthly lags (monthly); cross-model comparisons use the common sample shared by the candidate series, and a longer native monthly sample, where reported, is labeled separately and not used for the horse race.
	
	The monthly analysis is thus a frequency diagnostic, not a replacement for the daily design: a body--tail pattern that weakens, disappears, or reverses at the monthly frequency is informative in its own right, making internal pricing consistency not only model-dependent but frequency-dependent.
	
	\subsection{Body--Tail Test}
	\label{subsec:body-tail}
	
	The core test decomposes the full market portfolio into a body and a tail---tradable portfolios from the same investible universe as the aggregate market, yet distinct test assets. The reconstruction identity below does not impose zero alpha on either leg when the aggregate alpha is small; it provides a controlled way to examine how pricing errors distribute across a size-ranked decomposition of a market the models price in aggregate.
	
	At each formation date $\tau$ I sort investible stocks by market capitalization in descending order. For a split ratio $p$, the body holds stocks with cumulative share at or below $p$ and the tail holds the rest, over the grid
	\[
	p\in\{0.50,0.60,0.75,0.80,0.85,0.90,0.95,0.975,0.99\} .
	\]
	The grid spans balanced splits and cases that isolate a small-capitalization tail. Portfolios form at each end of June from the universe and ranks observed then and hold from the first trading day of July through the next June, close to standard Fama--French formation and insulated from high-frequency capitalization changes at the split boundary.
	
	Leg returns are value-weighted buy-and-hold returns with formation-date weights; dividends and delisting returns are treated as in the full market return, and within-leg cash flows are reinvested proportionally into surviving securities. Because the legs are annual buy-and-hold portfolios while factor returns come from external providers, a leg can drift from its formation weights within the year. Any alpha from this mechanical drift is shared by the matched random split in Section~\ref{subsec:random-split}, which uses the identical universe, schedule, holding rule, and reinvestment accounting, so a body--tail pattern the random split does not reproduce cannot be attributed to drift alone.
	
	Let $w^{B}_{p,\tau}$ and $w^{T}_{p,\tau}$ be the initial body and tail market-value shares ($w^{B}_{p,\tau}+w^{T}_{p,\tau}=1$). During the holding period these shares vary with realized returns and reinvested cash flows, so the daily reconstruction identity uses previous-day-end shares:
	\begin{equation}
		R^{e}_{M,d}
		=
		w^{B}_{p,d-1}R^{e}_{B,p,d}
		+
		w^{T}_{p,d-1}R^{e}_{T,p,d},
		\qquad
		w^{B}_{p,d-1}+w^{T}_{p,d-1}=1,
	\end{equation}
	where $R^{e}_{M,d}$ is the full investible market excess return and $R^{e}_{B,p,d}$ and $R^{e}_{T,p,d}$ are the body and tail excess returns. The identity makes the legs reconstruct the same aggregate market return without making them equivalent to it as test assets.
	
	I estimate leg pricing errors with standard time-series regressions,
	\begin{equation}
		R^{e}_{\ell,p,d}
		=
		\alpha_{\ell,p}
		+
		\beta_{\ell,p}'f_d
		+
		\varepsilon_{\ell,p,d},
		\qquad
		\ell\in\{B,T\},
	\end{equation}
	where $f_d$ is the daily factor vector; the monthly implementation estimates the same regression after compounding leg returns to calendar months and replacing $f_d$ with the monthly factor vector $f_m$:
	\begin{equation}
		R^{e}_{\ell,p,m}
		=
		\alpha_{\ell,p}^{(m)}
		+
		\left(\beta_{\ell,p}^{(m)}\right)'f_m
		+
		\varepsilon_{\ell,p,m}^{(m)} .
	\end{equation}
	Daily alphas are annualized by multiplying the intercept by 252, monthly alphas by 12.
	
	For each split ratio and model I test each leg alpha and the joint restriction
	\begin{equation}
		H_0:\alpha_{B,p}=\alpha_{T,p}=0
	\end{equation}
	with a Wald test using the between-leg residual covariance and Newey--West covariance matrices. Because the nine ratios are nested cuts of the same universe, their tests are not independent: I read rejection counts as within-model robustness and base cross-model statements on contrasts over the identical grid. I also estimate the recombined net-portfolio alpha and verify the reconstruction identity numerically, confirming that a small, insignificant aggregate alpha does not mechanically determine the leg alphas.
	
	\subsection{Random Split Test}
	\label{subsec:random-split}
	
	Random splits test whether the body--tail results are specific to size-ranked assignment rather than to the act of splitting. They use the same universe, formation dates, holding periods, split ratios, accounting, models, and test statistics as the body--tail test; only the assignment rule changes, with stocks assigned to two groups at random at each formation date.
	
	Each random split is matched to a body--tail ratio: at $p=0.80$, one leg holds about 80\% of market value and the other about 20\%, giving the random and body--tail tests the same aggregate portfolio and similar leg shares. At formation date $\tau$ I randomly permute the investible stocks and assign them to the A-leg until cumulative market capitalization reaches $p$, the remainder forming the B-leg. Each pair satisfies the same daily reconstruction identity in previous-day-end shares:
	\begin{equation}
		R^{e}_{M,d}
		=
		w^{A}_{p,d-1}R^{e}_{A,p,d}
		+
		w^{B}_{p,d-1}R^{e}_{B,p,d},
		\qquad
		w^{A}_{p,d-1}+w^{B}_{p,d-1}=1 .
	\end{equation}
	Monthly random-split results compound the same daily leg returns using the rule in Section~\ref{subsec:frequency-design}.
	
	The random split is not a structural simulation under the null that a model is true. It is a matched placebo for the split procedure, showing what alphas arise simply from dividing the same market into two value-weighted legs of similar share---which matters because the smaller leg grows more volatile and less precisely estimated as $p$ rises. The contrast is the main evidence: similar leg alphas and rejection rates across the deterministic and random designs would indict splitting itself, whereas small or directionless random-split results alongside systematic body--tail rejections point to the size-ranked assignment rule. I therefore treat the random split as a matched placebo, not a secondary robustness check.
	
	\section{Spanning Tests Across Factor Models}
	\label{sec:span}
	
	Before the body--tail test, I examine mean--variance inclusion relations among the traded factor models. As Section~\ref{subsec:two-criteria} notes, \citet{BS17} establish that for traded models the comparison of one against another is governed by factor spanning and linked to the maximum Sharpe ratio, so common test-asset alphas do not determine the ranking. Running the spanning tests first fixes which model is favored on that criterion, so any later body--tail evidence is read against that baseline rather than as a generic weakness. The two analyses answer different questions: a model that leaves significant body--tail alphas is not thereby inferior in the mean--variance sense, and a larger opportunity set need not imply pricing of the market's internal components.
	
	\subsection{Test Design}
	
	FF6 consists of the market, size, value, profitability, investment, and momentum factors; q5 of the market, size, investment, return-on-equity, and expected-growth factors. Portfolio pricing tests use each provider's market factor and risk-free rate, but the mutual spanning tests use a common market factor, to avoid placing two nearly identical market factors on opposite sides of the same regression. Baseline results use the Fama--French market factor; I confirm them with the global-q market factor. I define the nonmarket blocks as
	\begin{align*}
		g_t^{q}
		&=
		(ME_t,IA_t,ROE_t,EG_t)^{\prime},
		\\
		g_t^{FF}
		&=
		(SMB_t,HML_t,RMW_t,CMA_t,MOM_t)^{\prime},
	\end{align*}
	and estimate the mutual spanning regressions
	\begin{align}
		g_t^{q}
		&=
		\boldsymbol{\alpha}_{q\mid FF6}
		+
		B_{q\mid FF6}f_t^{FF6}
		+
		\boldsymbol{u}_{q,t},
		\label{eq:span_ff6_q5}
		\\
		g_t^{FF}
		&=
		\boldsymbol{\alpha}_{FF\mid q5}
		+
		B_{FF\mid q5}f_t^{q5}
		+
		\boldsymbol{u}_{FF,t},
		\label{eq:span_q5_ff6}
	\end{align}
	with nulls $H_0:\boldsymbol{\alpha}_{q\mid FF6}=\boldsymbol{0}$ and $H_0:\boldsymbol{\alpha}_{FF\mid q5}=\boldsymbol{0}$. I test the joint intercepts with the GRS test and a HAC Wald test (Newey--West lag 21 trading days for daily data, 6 months for monthly). The common daily sample is January 3, 1967 to December 31, 2024 (14,598 observations); the monthly sample, from compounding daily returns within calendar months, has 696. I measure the excluded block's economic contribution by
	\begin{equation}
		\Delta\operatorname{SR}^{2}
		=
		\widehat{\boldsymbol{\alpha}}^{\prime}
		\widehat{\Sigma}_{u}^{-1}
		\widehat{\boldsymbol{\alpha}},
		\label{eq:span_delta_sr2}
	\end{equation}
	the annualized increase in the squared maximum Sharpe ratio; the spanning decision itself rests on the joint intercept tests.
	
	\subsection{Joint and Factor-Level Spanning Results}
	
	Table~\ref{tab:span-joint} reports the joint results. In daily data both nulls are strongly rejected---neither model fully absorbs the other's nonmarket factors---but the magnitudes are asymmetric: adding the q5 block to FF6 raises the annualized maximum Sharpe ratio from 1.311 to 2.205 ($\Delta\operatorname{SR}^{2}=3.143$), whereas adding the FF6 block to q5 raises it only from 2.077 to 2.205 ($\Delta\operatorname{SR}^{2}=0.549$).
	
	The monthly tests reject asymmetrically too. The null that FF6 spans q5 is strongly rejected, while in the reverse direction the GRS and HAC $p$-values straddle the 5\% level (0.051 and 0.044), so the decision depends on the inference method; the global-q market factor leaves this unchanged. Monthly data thus suggest q5 absorbs much of the FF6 block without supporting robust complete spanning.
	
	\FloatBarrier
	\begin{table}[H]
		\centering
		\singlespacing
		\caption{Joint Spanning Tests Across Factor Models and Maximum Sharpe Ratios}
		\label{tab:span-joint}
		\footnotesize
		\setlength{\tabcolsep}{4.2pt}
		\begin{tabular}{llrrrrrr}
			\toprule
			Frequency & Null hypothesis & GRS & GRS $p$ & HAC Wald & HAC $p$
			& $\operatorname{SR}_{0}$ & $\Delta\operatorname{SR}^{2}$ \\
			\midrule
			Daily   & FF6 $\supseteq$ q5  & 45.197 & $<0.001$ & 148.177 & $<0.001$ & 1.311 & 3.143 \\
			Daily   & q5 $\supseteq$ FF6  &  6.248 & $<0.001$ &  23.086 & $<0.001$ & 2.077 & 0.549 \\
			Monthly & FF6 $\supseteq$ q5  & 34.500 & $<0.001$ & 114.540 & $<0.001$ & 1.222 & 2.687 \\
			Monthly & q5 $\supseteq$ FF6  &  2.217 & 0.051    &  11.426 & 0.044    & 1.982 & 0.255 \\
			\bottomrule
		\end{tabular}
		
		\vspace{0.4em}
		\parbox{0.96\textwidth}{\footnotesize
			Notes: $FF6\supseteq q5$ is the null that FF6 jointly spans the q5-specific factors ME, IA, ROE, and EG. $q5\supseteq FF6$ is the null that q5 jointly spans the FF6-specific factors SMB, HML, RMW, CMA, and MOM. $\operatorname{SR}_{0}$ is the annualized sample maximum Sharpe ratio of the benchmark model. $\Delta\operatorname{SR}^{2}$ is the annualized increase in the squared maximum Sharpe ratio from adding the other model's factor block. The reported results use the Fama--French market factor as the common market factor. The spanning decisions are unchanged when I use the global-q market factor. The HAC lag is 21 for daily data and 6 for monthly data.}
	\end{table}
	\FloatBarrier
	
	Table~\ref{tab:span-factor-level} locates the asymmetry. FF6 almost fully absorbs ME and renders IA insignificant at 5\%, but ROE and EG leave significant annualized alphas of 313.6 and 864.4 bp; EG alone has an individual $\Delta\operatorname{SR}^{2}$ of 2.425, the single largest source of the q5 block's expansion of the FF6 opportunity set. In the reverse direction q5 largely absorbs HML, RMW, CMA, and MOM, leaving only SMB significant (92.3 bp), which drives the borderline monthly joint test. Because the joint statistic uses the residual covariance matrix, individual $\Delta\operatorname{SR}^{2}$ values should not be summed.
	
	\FloatBarrier
	\begin{table}[H]
		\centering
		\singlespacing
		\caption{Monthly Factor-Level Spanning Regressions}
		\label{tab:span-factor-level}
		\footnotesize
		\setlength{\tabcolsep}{5.2pt}
		\begin{tabular}{llrrrr}
			\toprule
			Benchmark model & Target factor & Annualized alpha (bp) & HAC $p$ & $R^{2}$
			& Individual $\Delta\operatorname{SR}^{2}$ \\
			\midrule
			\multicolumn{6}{l}{\textit{Panel A: Pricing q5-specific factors with FF6}} \\
			FF6 & ME  &  17.6 & 0.605    & 0.950 & 0.005 \\
			FF6 & IA  &  74.6 & 0.070    & 0.855 & 0.074 \\
			FF6 & ROE & 313.6 & $<0.001$ & 0.642 & 0.347 \\
			FF6 & EG  & 864.4 & $<0.001$ & 0.449 & 2.425 \\
			\addlinespace
			\multicolumn{6}{l}{\textit{Panel B: Pricing FF6-specific factors with q5}} \\
			q5 & SMB &  92.3 & 0.006 & 0.956 & 0.170 \\
			q5 & HML & 119.8 & 0.348 & 0.485 & 0.025 \\
			q5 & RMW &  13.3 & 0.895 & 0.473 & 0.001 \\
			q5 & CMA & -13.7 & 0.755 & 0.861 & 0.003 \\
			q5 & MOM & -42.0 & 0.843 & 0.277 & 0.001 \\
			\bottomrule
		\end{tabular}
		
		\vspace{0.4em}
		\parbox{0.96\textwidth}{\footnotesize
			Notes: The sample covers 696 months from January 1967 to December 2024. Alphas are monthly intercepts multiplied by 12 and reported in basis points. HAC $p$-values use a Newey--West lag of 6 months. Individual $\Delta\operatorname{SR}^{2}$ is the annualized increase in the squared maximum Sharpe ratio from adding each target factor alone to the benchmark model.}
	\end{table}
	\FloatBarrier
	
	The spanning tests therefore give q5 a favorable baseline: ROE and EG expand the opportunity set beyond FF6 by economically large amounts, and in monthly data q5 absorbs most of the FF6 block. The body--tail evidence below should not be read as q5 being inferior in the mean--variance sense; the narrower point the paper develops is that a model can be strong on the criterion of \citet{BS17} and still leave systematic alphas on a size-ranked decomposition of the market it prices. It is worth flagging which factors carry that tension: the same ROE and expected-growth block that most expands q5's opportunity set here is the block that Section~\ref{sec:diagnostics} traces the body--tail pattern to. The factors that most enlarge the frontier are thus the ones most associated with the leg-level errors---exactly the separation between the two criteria that an approximate SDF permits.
	
	\section{Daily Full--Period Empirical Results}
	\label{sec:daily-fullperiod}
	
	This section reports the full-period evidence at the daily frequency, the native frequency at which the investible universe, market return, body--tail legs, and matched random splits are constructed and at which the baseline regressions use the providers' daily factor series. The results should therefore be read as daily-frequency evidence; whether the same model ranking survives monthly aggregation is taken up separately in Section~\ref{sec:monthly}. The market-level sanity check in Section~\ref{sec:data} establishes that all candidate models price the production market portfolio closely at the daily frequency, so the tests below do not begin from an aggregate return the models already fail to price. They ask a narrower question: when the same daily universe is organized into size-ranked body and tail portfolios, do additional alpha patterns appear, and do they differ across models that all pass the aggregate daily benchmark?
	
	\subsection{Daily Body--Tail Test}
	\label{subsec:daily-body-tail}
	
	Table~\ref{tab:body_tail_full} summarizes the daily full-period body--tail tests. For each split ratio $p$, the body holds the top $p$ share of cumulative market capitalization at formation and the tail the rest; for each pair I estimate daily time-series alpha regressions and test the joint zero-alpha null with a Wald test, using Newey--West standard errors with 21 daily lags as in Section~\ref{subsec:body-tail}.
	
	At the daily frequency the models separate sharply on the identical split grid. CAPM, Carhart, FF5, and FF6 reject the joint null at no ratio under the baseline HAC lag. FF3 rejects at four ratios, though its reconstructed net-market alpha stays insignificant. q5 is the daily outlier, rejecting at all nine ratios with a negative average body alpha and a positive average tail alpha. As noted in Section~\ref{subsec:body-tail}, the nine ratios are nested cuts of the same universe, so the counts describe within-model robustness, not nine independent tests. The informative object is the cross-model contrast at the same frequency: under one reconstruction identity, universe, split grid, and statistic, q5 produces a systematic opposite-signed leg pattern---and FF3 a weaker mirror-image one---that the aggregate daily benchmark does not reveal.
	
	\FloatBarrier
	\begin{table}[H]
		\centering
		\singlespacing
		\caption{Daily full-period body--tail tests}
		\label{tab:body_tail_full}
		\footnotesize
		\begin{tabular}{lrrrrrr}
			\toprule
			Model & Joint reject & Alpha-sum reject & Net reject & Body $\alpha$ & Tail $\alpha$ & Median joint $p$ \\
			& $(/9)$ & $(/9)$ & $(/9)$ & (bp/yr) & (bp/yr) & \\
			\midrule
			CAPM    & 0/9 & 0/9 & 0/9 & -18.3 &   82.3 & 0.775 \\
			FF3     & 4/9 & 4/9 & 0/9 &  18.5 & -101.6 & 0.212 \\
			Carhart & 0/9 & 0/9 & 0/9 &   7.0 &  -32.0 & 0.612 \\
			FF5     & 0/9 & 0/9 & 0/9 &  -3.1 &  -13.6 & 0.721 \\
			FF6     & 0/9 & 0/9 & 0/9 & -10.7 &   36.2 & 0.588 \\
			q5      & 9/9 & 8/9 & 0/9 & -56.2 &  181.0 & $6.25 \times 10^{-5}$ \\
			\bottomrule
		\end{tabular}
		
		\vspace{0.4em}
		\parbox{0.95\textwidth}{\footnotesize
			\emph{Notes:} The table summarizes daily body--tail pair tests across nine split ratios, $p \in \{0.50,0.60,0.75,0.80,0.85,0.90,0.95,0.975,0.99\}$. The sample is the common daily sample, 1967--2024, with Newey--West 21 daily lags. ``Joint reject'' counts rejections of the two-leg zero-alpha Wald test at the 5\% level. ``Alpha-sum reject'' counts rejections for the unweighted sum of the two leg alphas. ``Net reject'' counts rejections for the value-weighted net market portfolio reconstructed from the two legs. Because the nine ratios are nested, the counts describe within-model robustness, not independent tests.}
	\end{table}
	\FloatBarrier
	
	Table~\ref{tab:q5_body_tail_by_ratio} reports the daily q5 results by split ratio. The body alpha is negative and the tail alpha positive at every ratio, so the sign pattern is not generated by one cutoff: it holds from the balanced split to the small-tail splits, with the tail alpha staying positive even when the tail leg is small, though precision and magnitude vary with $p$.
	
	\FloatBarrier
	\begin{table}[H]
		\centering
		\singlespacing
		\begin{threeparttable}
			\caption{Daily q5 body--tail alphas by split ratio}
			\label{tab:q5_body_tail_by_ratio}
			\footnotesize
			\begin{tabular}{rrrrrrrr}
				\toprule
				Split $p$ & Body $\alpha$ & Tail $\alpha$ & $t_A$ & $t_B$ & Joint $p$ & $\alpha_A+\alpha_B$ & Sum $p$ \\
				& (bp/yr) & (bp/yr) & & & & (bp/yr) & \\
				\midrule
				0.500 & -158.4 & 147.9 & -4.54 & 4.37 & $3.30 \times 10^{-5}$ & -10.5 & 0.191 \\
				0.600 & -119.4 & 165.4 & -4.60 & 4.47 & $2.57 \times 10^{-5}$ & 45.9 & $7.40 \times 10^{-4}$ \\
				0.750 &  -74.6 & 201.1 & -5.01 & 5.00 & $2.40 \times 10^{-6}$ & 126.5 & $2.47 \times 10^{-6}$ \\
				0.800 &  -59.1 & 207.8 & -5.03 & 5.14 & $1.21 \times 10^{-6}$ & 148.7 & $8.36 \times 10^{-7}$ \\
				0.850 &  -38.0 & 177.1 & -4.17 & 4.29 & $6.25 \times 10^{-5}$ & 139.1 & $3.82 \times 10^{-5}$ \\
				0.900 &  -23.8 & 156.5 & -3.37 & 3.53 & 0.001 & 132.7 & $6.75 \times 10^{-4}$ \\
				0.950 &  -14.4 & 159.1 & -2.71 & 2.94 & 0.007 & 144.6 & 0.005 \\
				0.975 &   -9.8 & 152.0 & -2.11 & 2.23 & 0.040 & 142.2 & 0.031 \\
				0.990 &   -8.5 & 262.0 & -1.98 & 2.90 & 0.009 & 253.5 & 0.004 \\
				\bottomrule
			\end{tabular}
			\begin{tablenotes}[flushleft]
				\footnotesize
				\item Notes: This table reports the daily q5 body--tail results for each split ratio in the common daily sample with Newey--West 21 daily lags. $A$ denotes the body leg and $B$ denotes the tail leg. Alphas are annualized basis points. The net reconstructed market alpha is -5.8 bp per year with $p=0.151$ for every split ratio because the two legs reconstruct the same production market return.
			\end{tablenotes}
		\end{threeparttable}
	\end{table}
	\FloatBarrier
	
	The daily leg pattern is invisible in the aggregate benchmark. Recombining the body and tail legs with their value weights recovers the production market portfolio, and the net alpha is insignificant for all models, including q5 (Table~\ref{tab:q5_body_tail_by_ratio}, net $\alpha=-5.8$ bp per year, $p=0.151$). This is not a contradiction but the diagnostic content of the exercise: the recombination identity holds for every model, so it cannot by itself produce a leg pattern in one model and not in others. The daily q5 result is therefore a decomposition-specific pattern across size-ranked components---not a failure to price the aggregate daily market, and not a mechanical consequence of splitting a portfolio into two legs.
	
	\subsection{Random Split as a Matched Placebo-Style Benchmark}
	\label{subsec:daily-random-split-results}
	
	The body--tail design holds the reconstruction identity fixed but cannot, by itself, separate the size-ranked assignment rule from the act of splitting; the random split does so approximately. As described in Section~\ref{subsec:random-split}, it shares the universe, target split ratios, aggregate market portfolio, sample, model, statistic, formation schedule, holding rule, and reinvestment accounting with the daily body--tail test, changing only the assignment rule. The match is not exact in realized cap shares---market capitalization is highly right-skewed, so random ordering cannot hit a target cumulative share $p$ without splitting individual stocks---so I read it as an approximately matched placebo rather than a one-for-one replica. As Section~\ref{subsec:random-split} sets out, it absorbs the first-order effects of dividing the same market into a large and a small leg at comparable ratios---including the rising volatility and estimation error of the shrinking tail---leaving the size-ranked rule as the operative difference: similar leg alphas and rejection rates under both designs would indict splitting itself, whereas weak or directionless random splits alongside systematic body--tail rejections point to the size-ranked rule.
	
	Table~\ref{tab:random_split_full} makes this comparison at the daily frequency. For q5, the random-split joint rejection rate is 4.1\%, near the nominal 5\% level: when the same daily market is split at comparable target ratios without size ranking, the q5 pattern collapses to noise, while size-ranked splitting rejects at all nine ratios. The daily q5 result is therefore concentrated in the size-ordered decomposition, not a generic split effect.
	
	The contrast is also model-specific, but not in a way that isolates q5 alone. For CAPM, Carhart, FF5, and FF6, the random-split rejection rate \emph{exceeds} the deterministic body--tail rate, so size ranking is stabilizing: disciplined size-ranked splitting rejects less often than random splitting at matched ratios. Only q5 and FF3 reverse this---deterministic ranking rejects more than the placebo---so size ranking is destabilizing for both. The two differ in kind. q5 rejects at all nine ratios with a body-negative, tail-positive pattern against a 4.1\% placebo rate, whereas FF3 rejects at four ratios with the mirror-image body-positive, tail-negative pattern against a much higher 13.2\% placebo rate. At the daily frequency q5 is thus the sharply destabilized case and FF3 the milder one---an ordering that Section~\ref{sec:monthly} shows is not frequency-invariant.
	
	\FloatBarrier
	\begin{table}[H]
		\centering
		\singlespacing
		\begin{threeparttable}
			\caption{Daily body--tail tests versus random-split benchmarks}
			\label{tab:random_split_full}
			\footnotesize
			\begin{tabular}{lrrrrrr}
				\toprule
				Model & BT joint & Random joint & BT sum & Random sum & BT net & Random net \\
				& reject $(/9)$ & reject (\%) & reject $(/9)$ & reject (\%) & reject $(/9)$ & reject (\%) \\
				\midrule
				CAPM    & 0/9 &  2.3 & 0/9 &  5.3 & 0/9 & 0.0 \\
				FF3     & 4/9 & 13.2 & 4/9 & 18.0 & 0/9 & 0.0 \\
				Carhart & 0/9 &  8.6 & 0/9 & 13.2 & 0/9 & 0.0 \\
				FF5     & 0/9 & 15.6 & 0/9 & 20.0 & 0/9 & 0.0 \\
				FF6     & 0/9 & 12.6 & 0/9 & 17.3 & 0/9 & 0.0 \\
				q5      & 9/9 &  4.1 & 8/9 &  5.1 & 0/9 & 0.0 \\
				\bottomrule
			\end{tabular}
			\begin{tablenotes}[flushleft]
				\footnotesize
				\item Notes: ``BT'' denotes the deterministic body--tail split. Random-split rejection rates are weighted averages across the same nine target split ratios, with 500 random pairs per ratio and model. Because market capitalization is highly right-skewed and individual stocks are not split, the realized random-leg market shares need not equal the target ratios exactly. The random split should therefore be read as an approximately matched placebo-style benchmark. All tests use the common daily sample and Newey--West 21 daily lags. The random split preserves the same investible universe, target split ratio, net market portfolio, factor model, sample period, and test statistic; only the size-ranked assignment rule is removed.
			\end{tablenotes}
		\end{threeparttable}
	\end{table}
	\FloatBarrier
	
	Table~\ref{tab:q5_random_by_ratio} gives the same q5 comparison by ratio. The deterministic daily body--tail joint test rejects at every ratio, while the matched random-split rate stays low throughout: even at $p=0.99$, where the tail leg is very small, the random joint rejection rate is only 8.2\%, so the q5 result is not produced mechanically by a small tail leg but concentrated in the size-ranked tail. The random-split alpha magnitudes rise smoothly with the target ratio ($|\alpha_B|$ from 19.1 to 152.5 bp per year), tracing the estimation-noise channel the placebo is designed to capture, and the deterministic q5 tail alphas in Table~\ref{tab:q5_body_tail_by_ratio} sit beyond that benchmark.
	
	\FloatBarrier
	\begin{table}[H]
		\centering
		\singlespacing
		\resizebox{\textwidth}{!}{
			\begin{threeparttable}
				\caption{Daily matched random-split benchmark for q5}
				\label{tab:q5_random_by_ratio}
				\footnotesize
				\begin{tabular}{rrrrrr}
					\toprule
					Split $p$ & BT joint $p$ & Random joint reject & Random sum reject & Random $|\alpha_A|$ & Random $|\alpha_B|$ \\
					& & (\%) & (\%) & (bp/yr) & (bp/yr) \\
					\midrule
					0.500 & $3.30 \times 10^{-5}$ & 3.0 & 0.0 & 18.9 & 19.1 \\
					0.600 & $2.57 \times 10^{-5}$ & 1.4 & 2.0 & 15.4 & 22.3 \\
					0.750 & $2.40 \times 10^{-6}$ & 2.8 & 3.8 & 10.7 & 30.2 \\
					0.800 & $1.21 \times 10^{-6}$ & 3.4 & 4.6 & 9.9 & 36.8 \\
					0.850 & $6.25 \times 10^{-5}$ & 2.8 & 5.0 & 8.9 & 43.3 \\
					0.900 & 0.001 & 3.8 & 5.8 & 7.6 & 52.9 \\
					0.950 & 0.007 & 5.4 & 8.0 & 6.6 & 77.0 \\
					0.975 & 0.040 & 6.0 & 7.8 & 6.1 & 104.9 \\
					0.990 & 0.009 & 8.2 & 8.6 & 5.8 & 152.5 \\
					\bottomrule
				\end{tabular}
				\begin{tablenotes}[flushleft]
					\footnotesize
					\item Notes: For each target split ratio, the table compares the deterministic daily q5 body--tail joint-test $p$-value with the empirical rejection rate from 500 random splits. Because the market-capitalization distribution is highly skewed, random splits are approximately matched to the target ratio rather than exact cap-share replicas. All tests use Newey--West 21 daily lags. Random-split alpha magnitudes are mean absolute annualized alphas across the 500 random pairs.
				\end{tablenotes}
			\end{threeparttable}
		}
	\end{table}
	\FloatBarrier
	
	
	\section{Daily Subperiod Empirical Results}
	\label{sec:subperiod}
	
	The full-period daily evidence shows a sharp body--tail separation for q5. I next ask where in the sample this daily-frequency pattern concentrates, examining three complementary time decompositions: calendar subperiods, rolling 10-year windows, and regimes detected from the q5 rolling rejection signal. The analysis stays conditional on the daily implementation---daily buy-and-hold legs, daily factor returns, and daily Newey--West standard errors with 21 trading-day lags throughout---so the aim is to describe when the daily pattern is active, not to make a frequency-invariant statement.
	
	\subsection{Calendar Subperiods}
	
	Table~\ref{tab:subperiod_decade} reports, for each model and calendar period, how many of the nine split ratios reject the joint zero-alpha null at the 5\% level. The first row begins in January 1967 and the last ends in December 2024.
	
	\FloatBarrier
	\begin{table}[H]
		\centering
		\singlespacing
		\begin{threeparttable}
			\caption{Calendar subperiod body--tail rejections: daily tests}
			\label{tab:subperiod_decade}
			\footnotesize
			\begin{tabular}{lrrrrrrr}
				\toprule
				Period & Obs. & CAPM & FF3 & Carhart & FF5 & FF6 & q5 \\
				\midrule
				1967--1969 & 727 & 0/9 & 0/9 & 0/9 & 0/9 & 1/9 & \textbf{4/9} \\
				1970--1979 & 2,526 & 0/9 & 0/9 & 0/9 & 0/9 & 0/9 & \textbf{2/9} \\
				1980--1989 & 2,528 & 0/9 & 3/9 & 2/9 & 5/9 & 5/9 & \textbf{9/9} \\
				1990--1999 & 2,528 & 0/9 & 4/9 & 0/9 & 0/9 & 0/9 & 0/9 \\
				2000--2009 & 2,515 & 4/9 & 0/9 & 1/9 & 2/9 & 3/9 & \textbf{4/9} \\
				2010--2019 & 2,516 & 0/9 & 0/9 & 0/9 & 0/9 & 0/9 & 0/9 \\
				2020--2024 & 1,258 & 0/9 & 3/9 & 3/9 & 0/9 & 0/9 & 0/9 \\
				\bottomrule
			\end{tabular}
			\begin{tablenotes}[flushleft]
				\footnotesize
				\item Notes: Each entry reports the number of rejected joint zero-alpha tests among the nine body--tail split ratios. The sample is the common daily sample. HAC standard errors use 21 daily lags. Bold entries highlight q5 rejections.
			\end{tablenotes}
		\end{threeparttable}
	\end{table}
	\FloatBarrier
	
	The daily rejection pattern is not uniform over time. q5 rejects at all nine split ratios in the 1980s, at some ratios in the late 1960s, 1970s, and 2000s, and at none in the 1990s, the 2010s, or 2020--2024. The full-period daily q5 result is therefore neither a single-period artifact nor a time-invariant failure.
	
	Other models also reject in some subperiods. FF3 rejects at four split ratios in the 1990s and three in 2020--2024; FF5 and FF6 reject in the 1980s; CAPM rejects in 2000--2009. These episodes matter because they show the diagnostic is not built to single out one model mechanically. Even so, q5 has the highest rejection frequency across the seven subperiods---30.2\% of its model--ratio cells, against 6.3\% (CAPM) to 15.9\% (FF3) for the others---where each share is the rejections summed over the seven subperiods, unweighted, divided by the 63 model--ratio cells.
	
	The interpretation is therefore conditional and comparative: at the daily frequency, q5 is the model whose body--tail pattern is most persistent across calendar subperiods.
	
	\subsection{Rolling 10-Year Windows}
	
	Calendar boundaries are arbitrary. I therefore repeat the daily body--tail test in rolling 10-year windows, which smooth short-window noise while allowing the active periods to be located more flexibly. Table~\ref{tab:rolling10_summary} summarizes 600 rolling windows with at least 2,000 common-sample daily observations.
	
	\FloatBarrier
	\begin{table}[H]
		\centering
		\singlespacing
		\begin{threeparttable}
			\caption{Rolling 10-year body--tail rejections: daily tests}
			\label{tab:rolling10_summary}
			\footnotesize
			\begin{tabular}{lrrrr}
				\toprule
				Model & Windows & Mean reject (\%) & Median reject (\%) & Any reject (\%) \\
				\midrule
				CAPM & 600 & 8.1 & 0.0 & 15.0 \\
				FF3 & 600 & 14.4 & 0.0 & 38.3 \\
				Carhart & 600 & 9.4 & 0.0 & 27.8 \\
				FF5 & 600 & 13.6 & 0.0 & 29.0 \\
				FF6 & 600 & 14.1 & 0.0 & 30.2 \\
				\textbf{q5} & 600 & \textbf{40.1} & \textbf{44.4} & \textbf{79.0} \\
				\bottomrule
			\end{tabular}
			\begin{tablenotes}[flushleft]
				\footnotesize
				\item Notes: The table summarizes 10-year rolling windows with at least 2,000 common-sample daily observations. Mean and median rejection shares are computed across the nine body--tail split ratios within each window. ``Any rejection'' is the share of rolling windows in which at least one split ratio rejects joint zero alpha at the 5\% level. HAC standard errors use 21 daily lags.
			\end{tablenotes}
		\end{threeparttable}
	\end{table}
	\FloatBarrier
	
	The rolling windows sharpen the daily-frequency contrast. q5 rejects 40.1\% of split ratios on average, has a median rejection share of 44.4\%, and has at least one rejection in 79.0\% of rolling windows. The other five models have mean rejection shares between 8.1\% and 14.4\%, and their median rejection share is zero. Thus, within the daily rolling-window design, q5 is persistently the most rejection-prone model.
	
	Table~\ref{tab:rolling10_q5_by_end_decade} groups the q5 rolling windows by ending decade. The daily q5 pattern is strongest in windows ending in the 1980s and 1990s, weakens through the 2000s and 2010s, and nearly vanishes in windows ending in the 2020s.
	
	\FloatBarrier
	\begin{table}[H]
		\centering
		\singlespacing
		\begin{threeparttable}
			\caption{q5 rolling 10-year rejections by ending decade: daily tests}
			\label{tab:rolling10_q5_by_end_decade}
			\footnotesize
			\begin{tabular}{lrrrr}
				\toprule
				Ending decade & Windows & Mean reject (\%) & Median reject (\%) & Any reject (\%) \\
				\midrule
				1970s & 60 & 24.3 & 22.2 & 93.3 \\
				1980s & 120 & 57.4 & 66.7 & 84.2 \\
				1990s & 120 & 61.5 & 66.7 & 99.2 \\
				2000s & 120 & 38.2 & 44.4 & 87.5 \\
				2010s & 120 & 30.3 & 22.2 & 70.8 \\
				2020s & 60 & 2.0 & 0.0 & 13.3 \\
				\bottomrule
			\end{tabular}
			\begin{tablenotes}[flushleft]
				\footnotesize
				\item Notes: The table groups q5 rolling 10-year windows by the decade in which the rolling window ends. Rejection shares are computed across the nine body--tail split ratios. HAC standard errors use 21 daily lags.
			\end{tablenotes}
		\end{threeparttable}
	\end{table}
	\FloatBarrier
	
	This time profile is important for interpretation. A purely mechanical split artifact would be expected to appear more uniformly across periods. Instead, the daily q5 rejection signal switches on and off across the sample. This does not by itself identify the economic source of the pattern, but it motivates the factor-block diagnostics in Section~\ref{sec:diagnostics}, which ask whether the active daily periods line up with q5's nonmarket profitability--growth block.
	
	\subsection{Detected q5 Regimes}
	
	As a final daily-frequency diagnostic, I use the q5 rolling rejection signal to partition the sample into periods in which q5 body--tail rejections concentrate and periods in which they are absent. This partition is descriptive. It is not an out-of-sample test, a formal structural-break procedure, or an independent validation exercise, because the regimes are detected from the q5 rejection signal itself.
	
	\FloatBarrier
	\begin{table}[H]
		\centering
		\singlespacing
		\begin{threeparttable}
			\caption{Detected q5 regimes and model-level rejections: daily tests}
			\label{tab:q5_detected_regimes}
			\footnotesize
			\begin{tabular}{lrrrrrrr}
				\toprule
				Detected regime & Obs. & CAPM & FF3 & Carhart & FF5 & FF6 & q5 \\
				\midrule
				Nonreject 1967--1976 & 2,221 & 0/9 & 0/9 & 0/9 & 0/9 & 0/9 & 0/9 \\
				Reject 1976--1978 & 484 & 9/9 & 5/9 & 5/9 & 5/9 & 5/9 & \textbf{8/9} \\
				Nonreject 1978--1982 & 1,116 & 0/9 & 0/9 & 0/9 & 0/9 & 0/9 & 0/9 \\
				Reject 1982--1996 & 3,392 & 0/9 & 3/9 & 2/9 & 3/9 & 4/9 & \textbf{8/9} \\
				Nonreject 1996--2001 & 1,325 & 0/9 & 1/9 & 0/9 & 0/9 & 0/9 & 0/9 \\
				Reject 2001--2006 & 1,130 & 5/9 & 0/9 & 0/9 & 3/9 & 3/9 & \textbf{7/9} \\
				Nonreject 2006--2024 & 4,761 & 0/9 & 0/9 & 0/9 & 0/9 & 0/9 & 0/9 \\
				\bottomrule
			\end{tabular}
			\begin{tablenotes}[flushleft]
				\footnotesize
				\item Notes: Entries report the number of rejected joint zero-alpha tests among the nine body--tail split ratios. The initial partial transition segment in 1967 contains only 169 common-sample observations and is omitted. Regimes are detected from the q5 rolling rejection signal and are used only as a descriptive chronology. HAC standard errors use 21 daily lags.
			\end{tablenotes}
		\end{threeparttable}
	\end{table}
	\FloatBarrier
	
	Table~\ref{tab:q5_detected_regimes} reinforces the same message. In the nonreject regimes, every model passes at all split ratios, including q5. The partition therefore does not describe q5 as a chronic daily failure. It isolates periods in which the daily body--tail signal becomes active. In the 1982--1996 and 2001--2006 reject regimes, q5 rejects at 8/9 and 7/9 split ratios, respectively. Other models also reject in some active periods, but q5 is usually the most frequent and persistent rejection within the detected q5 regimes.
	
	The subperiod evidence supports two daily-frequency conclusions. First, the full-period q5 result is not generated by a single calendar segment: it appears in calendar subperiods, rolling 10-year windows, and detected active regimes. Second, the signal is strongly time-varying. It is most visible from the 1980s through the mid-2000s and much weaker in the recent sample. The next section uses this time profile as a clue for mechanism, while the later monthly analysis asks whether the same ranking survives after the return and factor data are aggregated to calendar months.
	
	
	\section{Monthly Empirical Results}
	\label{sec:monthly}
	
	The previous sections report the body--tail evidence at the daily frequency, the native frequency of the CRSP stock-return panel. I now repeat the same diagnostic after compounding the body and tail returns to calendar months and estimating the pricing regressions with monthly factor returns. This is not a secondary robustness check: it asks whether the internal pricing pattern documented at the daily frequency is invariant to the observation frequency at which the factor model is evaluated.
	
	The monthly design holds the economic decomposition fixed. The investible universe, June formation schedule, split ratios, buy-and-hold accounting, delisting treatment, and body--tail assignment rule are the same as in the daily tests. For each leg and split ratio I compound daily returns within each calendar month,
	\[
	R_{\ell,p,m}
	=
	\prod_{d\in m}\left(1+R_{\ell,p,d}\right)-1,
	\qquad
	\ell\in\{B,T\},
	\]
	and estimate the same time-series alpha regressions with monthly factor returns and the corresponding monthly risk-free rate. The common monthly sample runs from January 1967 to December 2024 and contains 696 months. Unless stated otherwise, monthly inference uses Newey--West standard errors with six monthly lags.
	
	\subsection{Monthly Body--Tail Test}
	
	Table~\ref{tab:monthly_body_tail_diagnostics} summarizes the monthly body--tail tests across the same nine split ratios used in the daily analysis, and the relative ranking changes. The q5 joint rejection count falls from nine of nine splits at the daily frequency to one of nine, while FF3 rejects the joint zero-alpha null in four of nine splits. The frequency change therefore does more than mechanically reduce power: it shifts the most visible joint-test weakness from q5 to FF3 in the common monthly sample.
	
	\FloatBarrier
	\begin{table}[H]
		\centering
		\singlespacing
		\begin{threeparttable}
			\caption{Monthly body--tail diagnostics across models}
			\label{tab:monthly_body_tail_diagnostics}
			\footnotesize
			\setlength{\tabcolsep}{4.4pt}
			\begin{tabular}{lrrrrrrr}
				\toprule
				Model & Joint reject & Sum reject & Net reject & Median $p_J$ & Median $p_S$ & Body $\alpha$ & Tail $\alpha$ \\
				& $(/9)$ & $(/9)$ & $(/9)$ & & & (bp/yr) & (bp/yr) \\
				\midrule
				CAPM & 0/9 & 0/9 & 0/9 & 0.966 & 0.873 & -2.6 & 7.1 \\
				FF3 & 4/9 & 4/9 & 0/9 & 0.168 & 0.077 & 20.5 & -105.5 \\
				Carhart & 0/9 & 0/9 & 0/9 & 0.633 & 0.612 & 7.8 & -29.2 \\
				FF5 & 0/9 & 0/9 & 0/9 & 0.591 & 0.322 & 12.3 & -49.9 \\
				FF6 & 0/9 & 0/9 & 0/9 & 0.820 & 0.655 & 1.9 & 12.4 \\
				q5 & 1/9 & 5/9 & 0/9 & 0.130 & 0.045 & -20.6 & 116.7 \\
				\bottomrule
			\end{tabular}
			\begin{tablenotes}[flushleft]
				\footnotesize
				\item Notes: The table summarizes monthly body--tail pair tests across nine split ratios, $p\in\{0.50,0.60,0.75,0.80,0.85,0.90,0.95,0.975,0.99\}$. The sample contains 696 months from January 1967 to December 2024. Alphas are monthly intercepts multiplied by 12 and reported in annual basis points. $p_J$ is the joint two-leg zero-alpha Wald-test $p$-value. $p_S$ is the $p$-value for $\alpha_A+\alpha_B=0$. HAC standard errors use six monthly lags. ``Net reject'' refers to the value-weighted net market portfolio reconstructed from the two legs.
			\end{tablenotes}
		\end{threeparttable}
	\end{table}
	\FloatBarrier
	
	The sign patterns also differ by model. FF3 carries a positive average body alpha and a negative average tail alpha, with four joint and four alpha-sum rejections. q5 keeps the daily sign direction---negative body, positive tail---but its joint evidence weakens sharply: the median joint $p$-value rises to 0.130 and only one split rejects. Yet q5 is not cleanly neutralized by aggregation. It still rejects the alpha-sum restriction in five of nine splits, holds the lowest median sum-test $p$-value among the models, and leaves an average tail alpha of 116.7 bp per year.
	
	\subsection{FF3 and q5 by Split Ratio}
	
	Table~\ref{tab:monthly_ff3_q5_by_ratio} reports the split-level results for the two models that drive the monthly comparison. FF3 rejects mainly in the upper-middle and upper-tail of the grid, with a positive body alpha and a negative tail alpha at every ratio. q5 shows the opposite sign pattern at every ratio but rejects the joint test only at $p=0.80$; its alpha-sum test is more persistent, rejecting at five ratios because the positive tail alpha dominates the smaller negative body alpha.
	
	\FloatBarrier
	\begin{table}[H]
		\centering
		\singlespacing
		\begin{threeparttable}
			\caption{Monthly FF3 and q5 body--tail alphas by split ratio}
			\label{tab:monthly_ff3_q5_by_ratio}
			\footnotesize
			\begin{tabular}{rrrrrrrrr}
				\toprule
				& \multicolumn{4}{c}{FF3} & \multicolumn{4}{c}{q5} \\
				\cmidrule(lr){2-5}\cmidrule(lr){6-9}
				Split $p$ & Body $\alpha$ & Tail $\alpha$ & Joint $p$ & Sum $p$ & Body $\alpha$ & Tail $\alpha$ & Joint $p$ & Sum $p$ \\
				& (bp/yr) & (bp/yr) & & & (bp/yr) & (bp/yr) & & \\
				\midrule
				0.500 & 64.3 & -64.3 & 0.168 & 0.995 & -51.5 & 52.7 & 0.322 & 0.862 \\
				0.600 & 36.7 & -53.8 & 0.377 & 0.197 & -45.3 & 70.3 & 0.197 & 0.073 \\
				0.750 & 21.1 & -61.7 & 0.274 & 0.114 & -26.4 & 83.1 & 0.117 & 0.039 \\
				0.800 & 17.5 & -68.0 & 0.204 & 0.077 & -26.2 & 109.2 & 0.033 & 0.010 \\
				0.850 & 18.2 & -101.6 & 0.032 & 0.009 & -13.8 & 83.8 & 0.127 & 0.045 \\
				0.900 & 13.9 & -123.9 & 0.008 & 0.002 & -10.1 & 98.3 & 0.104 & 0.035 \\
				0.950 & 7.7 & -150.9 & 0.006 & 0.001 & -6.2 & 129.9 & 0.158 & 0.055 \\
				0.975 & 4.2 & -181.7 & 0.010 & 0.002 & -3.7 & 158.3 & 0.259 & 0.100 \\
				0.990 & 0.7 & -143.4 & 0.222 & 0.085 & -2.6 & 265.1 & 0.130 & 0.044 \\
				\bottomrule
			\end{tabular}
			\begin{tablenotes}[flushleft]
				\footnotesize
				\item Notes: This table reports monthly body--tail results for FF3 and q5 at each split ratio. The sample contains 696 months from January 1967 to December 2024. Body and tail alphas are monthly intercepts multiplied by 12 and reported in annual basis points. The joint test is the two-leg zero-alpha Wald test. The sum test is for $\alpha_A+\alpha_B=0$. HAC standard errors use six monthly lags.
			\end{tablenotes}
		\end{threeparttable}
	\end{table}
	\FloatBarrier
	
	The contrast does not reduce to lost power. A pure sample-size effect---14,598 trading days to 696 months---would weaken every q5 statistic in step, leaving q5 the dominant rejection case with smaller $t$-statistics. That is not what happens. The q5 \emph{joint} test collapses (nine of nine to one of nine) while its \emph{alpha-sum} test does not (eight of nine to five of nine, the lowest median sum-test $p$-value among all models), so aggregation changes the \emph{shape} of the q5 signal---from a body-negative, tail-positive pair the joint test detects to a tail-dominated imbalance that survives mainly in the sum test---rather than scaling it down uniformly. FF3, by contrast, is essentially unchanged across frequencies, with four joint rejections at both: it does not newly deteriorate, but q5 descends to its level. The monthly result is therefore a change in the cross-model ranking and in the form of the q5 pattern, not a uniform power loss. This is why the monthly analysis belongs in the main text, and why it shows the body--tail diagnostic to be frequency-dependent, not only model-dependent.
	
	\subsection{Common-Sample and Native-Sample Interpretation}
	
	The cross-model comparison uses the common monthly sample because q5 begins in 1967 while the Fama--French models have longer native histories, so the common sample is the appropriate horse-race sample, consistent with the design in Section~\ref{subsec:frequency-design}. Native-sample results remain useful for model-specific historical diagnostics but should not be used to rank models against q5: in its longer native sample FF3 has no monthly joint rejections, against four in the common 1967--2024 sample. The same model thus reads differently across samples, so monthly body--tail performance is sample- as well as frequency-dependent and must be compared within a fixed common sample.
	
	The monthly results revise the daily-only narrative. The daily evidence makes q5 the sharpest body--tail outlier; the monthly evidence shows this is not frequency-invariant, as q5 becomes much less fragile while FF3 becomes the more visible joint-test rejection case. The conclusion is therefore not that one model universally fails the body--tail test, but that internal pricing consistency depends on both the factor model and the frequency at which it is evaluated.
	
	\section{Robustness and Diagnostics}
	\label{sec:diagnostics}
	
	This section checks whether the daily body--tail evidence is driven by implementation choices and then isolates the factor block behind the daily q5 pattern. The robustness checks---alternative HAC lags, a $cMKT$ market-factor substitution, and external size-decile tests---ask whether the daily result is an artifact of a particular inference window, market-factor implementation, or test-asset construction. The factor-block diagnostics---ablations and a loading--premium decomposition---then locate the daily q5 imbalance in its nonmarket profitability--growth block, the same block that Section~\ref{sec:span} found to drive q5's spanning advantage. The monthly evidence is treated separately in Section~\ref{sec:monthly}, where the model ranking changes under lower-frequency aggregation. Unless stated otherwise, the diagnostics in this section use the common daily sample and Newey--West standard errors with 21 daily lags.
	
	\subsection{HAC-lag sensitivity}
	
	Table~\ref{tab:q5_hac_sensitivity} reports q5 joint zero-alpha $p$-values for the full-period body--tail test under Newey--West lags of 5, 21, 63, and 252 trading days.
	
	\FloatBarrier
	\begin{table}[H]
		\centering
		\singlespacing
		\begin{threeparttable}
			\caption{q5 body--tail joint-test $p$-values across HAC lags}
			\label{tab:q5_hac_sensitivity}
			\footnotesize
			\begin{tabular}{lrrrr}
				\toprule
				Split $p$ & NW5 & NW21 & NW63 & NW252 \\
				\midrule
				0.500 & $ 1.53 \times 10^{-5} $ & $ 3.30 \times 10^{-5} $ & $ 5.22 \times 10^{-5} $ & $ 7.80 \times 10^{-6} $ \\
				0.600 & $ 7.63 \times 10^{-6} $ & $ 2.57 \times 10^{-5} $ & $ 5.90 \times 10^{-5} $ & $ 2.53 \times 10^{-5} $ \\
				0.750 & $ 4.28 \times 10^{-7} $ & $ 2.40 \times 10^{-6} $ & $ 3.62 \times 10^{-6} $ & $ 7.46 \times 10^{-7} $ \\
				0.800 & $ 2.11 \times 10^{-7} $ & $ 1.21 \times 10^{-6} $ & $ 1.72 \times 10^{-6} $ & $ 5.10 \times 10^{-7} $ \\
				0.850 & $ 1.40 \times 10^{-5} $ & $ 6.25 \times 10^{-5} $ & $ 8.41 \times 10^{-5} $ & $ 2.17 \times 10^{-5} $ \\
				0.900 & $ 3.33 \times 10^{-4} $ & 0.001 & 0.001 & $ 8.02 \times 10^{-4} $ \\
				0.950 & 0.003 & 0.007 & 0.007 & 0.007 \\
				0.975 & 0.026 & 0.040 & 0.035 & 0.038 \\
				0.990 & 0.004 & 0.009 & 0.010 & 0.018 \\
				\bottomrule
			\end{tabular}
			\begin{tablenotes}[flushleft]
				\footnotesize
				\item Notes: The table reports Wald-test $p$-values for the joint zero-alpha restriction on the body and tail legs. The common daily sample is 1967--2024.
			\end{tablenotes}
		\end{threeparttable}
	\end{table}
	\FloatBarrier
	
	The result is insensitive to this choice: the joint null rejects at all nine ratios under every lag, with the largest $p$-value across the grid below 0.041. Survival under windows as long as 252 days is also a first piece of evidence against a pure high-frequency-microstructure reading---a pattern generated by daily nonsynchronous trading would not persist through HAC windows of a trading year. The check does not identify the economic source, but removes HAC bandwidth as an explanation.
	
	\subsection{Replacing the market factor with cMKT}
	
	A direct concern is market-factor mismatch: the q-factor market factor is not exactly my production market return. I therefore replace the native market factor in q5 and FF6 with cMKT, the same return the body and tail legs reconstruct.
	
	\FloatBarrier
	\begin{table}[H]
		\centering
		\singlespacing
		\begin{threeparttable}
			\caption{Market-factor replacement and body--tail rejections}
			\label{tab:cmkt_robustness}
			\footnotesize
			\begin{tabular}{lrrrrrrr}
				\toprule
				Model & Sample & Obs. & Joint & Median $p_J$ & $\bar\alpha_A$ & $\bar\alpha_B$ & Sum \\
				&        &      & reject &              & (bp/yr) & (bp/yr) & reject \\
				\midrule
				FF6 & 1963--2024 & 15481 & 0/9 & 0.567 & -10.3 & 36.7 & 0/9 \\
				cMKT--FF6 & 1963--2024 & 15481 & 1/9 & 0.425 & -9.9 & 38.3 & 1/9 \\
				q5 & 1967--2024 & 14598 & 9/9 & $<0.001$ & -56.2 & 181.0 & 8/9 \\
				cMKT--q5 & 1967--2024 & 14598 & 9/9 & $<0.001$ & -50.7 & 187.7 & 9/9 \\
				\bottomrule
			\end{tabular}
			\begin{tablenotes}[flushleft]
				\footnotesize
				\item Notes: ``Joint reject'' counts 5\% Wald rejections across the nine body--tail split ratios. ``Sum reject'' counts rejections for $\alpha_A+\alpha_B=0$. Alphas are annualized basis points. HAC standard errors use 21 daily lags. Net-market $p$-values for cMKT variants are omitted because the net portfolio is spanned by construction.
			\end{tablenotes}
		\end{threeparttable}
	\end{table}
	\FloatBarrier
	
	Table~\ref{tab:cmkt_robustness} confirms that the pattern survives replacement: native q5 and cMKT--q5 both reject at all nine cutoffs, with body and tail alphas of similar sign and magnitude. This is the decisive form of the check, because cMKT is the exact return the legs reconstruct and is built from CRSP source data rather than a provider series. If the pattern came from a mismatch between the q5 market factor and the market portfolio being decomposed, aligning the two would remove it; it does not. The source is therefore not the market factor, and the diagnostics below place it in the nonmarket block.
	
	\subsection{External size-portfolio diagnostics}
	\label{sec:external_size_diagnostics}
	
	The body--tail legs are internal test assets, built from the same universe that reconstructs the market return. I next ask whether the same size-ranked pattern appears in standard external size portfolios: daily value-weighted size deciles from the Kenneth French Data Library and from Open Source Asset Pricing (OSAP), which follows the open-source framework of \citet{CZ22}. The point is not another battery of test assets but a comparison with familiar size sorts whose construction is independent of my decomposition. I report deciles only; terciles and quintiles are coarser versions of the same French sort, and the decile cut is also the most informative about where the q5 errors sit across the size distribution.
	
	\FloatBarrier
	\begin{table}[H]
		\centering
		\singlespacing
		\begin{threeparttable}
			\caption{Daily size-decile joint-alpha tests}
			\label{tab:daily_size_decile_joint_tests}
			\footnotesize
			\setlength{\tabcolsep}{5pt}
			\begin{tabular}{llrrrrr}
				\toprule
				Source & Model & $p_J$ & Joint & Individual & Mean $\alpha$ & Median $|\alpha|$ \\
				& & & reject & rejects & (bp/yr) & (bp/yr) \\
				\midrule
				French & CAPM--FF & 0.468 & 0 & 0/10 & 100.1 & 99.3 \\
				French & FF3 & 0.005 & 1 & 3/10 & -46.0 & 39.2 \\
				French & Carhart & 0.005 & 1 & 2/10 & -43.6 & 43.9 \\
				French & FF5 & 0.124 & 0 & 1/10 & 5.2 & 33.6 \\
				French & FF6 & 0.120 & 0 & 1/10 & 4.1 & 27.9 \\
				French & CAPM--q5 & 0.469 & 0 & 0/10 & 99.4 & 98.6 \\
				French & q5 & $ 9.80 \times 10^{-7} $ & 1 & 8/10 & 130.7 & 146.3 \\
				\hline
				OSAP & CAPM--FF & 0.370 & 0 & 0/10 & 86.0 & 88.5 \\
				OSAP & FF3 & $ 4.02 \times 10^{-6} $ & 1 & 4/10 & -68.5 & 106.0 \\
				OSAP & Carhart & $ 1.14 \times 10^{-5} $ & 1 & 4/10 & -89.7 & 79.0 \\
				OSAP & FF5 & $ 3.46 \times 10^{-4} $ & 1 & 2/10 & 18.4 & 99.3 \\
				OSAP & FF6 & $ 5.45 \times 10^{-4} $ & 1 & 2/10 & -2.6 & 90.7 \\
				OSAP & CAPM--q5 & 0.366 & 0 & 0/10 & 84.8 & 87.7 \\
				OSAP & q5 & $ 3.05 \times 10^{-4} $ & 1 & 5/10 & 180.7 & 127.3 \\
				\bottomrule
			\end{tabular}
			\begin{tablenotes}[flushleft]
				\footnotesize
				\item Notes: The table reports Wald-test $p$-values for the joint zero-alpha restriction across daily value-weighted size deciles. The sample is January 3, 1967 to December 31, 2024. HAC standard errors use 21 daily lags. ``Joint reject'' is an indicator for rejection at the 5\% level. ``Individual rejects'' counts single-asset alpha rejections at the 5\% level. CAPM--FF uses the Fama--French market factor and risk-free rate. CAPM--q5 uses the q5 market factor and risk-free rate without the nonmarket q5 factors. The OSAP rows use port01--port10; the long--short portfolio is excluded.
			\end{tablenotes}
		\end{threeparttable}
	\end{table}
	\FloatBarrier
	
	Table~\ref{tab:daily_size_decile_joint_tests} shows construction dependence in the binary rejection but a stable direction underneath it. The rejection itself is library-dependent: in French deciles FF5 and FF6 pass while q5 rejects with eight of ten individual alphas significant, whereas in OSAP deciles FF5, FF6, and q5 all reject, so a pass/fail reading does not isolate q5. The direction of the nonmarket block does. Read against the CAPM--q5 baseline---the q5 market factor and risk-free rate with the nonmarket factors removed---q5 is the only candidate whose nonmarket block \emph{enlarges} the size-decile alpha relative to its own market-only specification: it moves the OSAP mean from 84.8 to 180.7 bp and the French mean from 99.4 to 130.7 bp, while the Fama--French blocks pull the alpha \emph{toward} zero (FF6 takes the OSAP mean from 86.0 to $-2.6$ bp, and FF5/FF6 sit near 5 bp in French). The rejection is shared in OSAP; the mechanism producing it is not. Because CAPM--q5 itself does not reject in either library, the rejection is a property of the full q5 block, not of the q5 market factor---the same separation the paper documents internally, now in externally constructed portfolios that cross-validate it. French, OSAP, and the body--tail legs are three independent ways to impose a size ranking on U.S. equities: the ranking of q5 against FF5 and FF6 shifts across them, but the direction of the q5 block---adding it widens rather than closes the size-ranked error---does not. Construction dependence is therefore not a reason to discount the body--tail result but additional evidence that it is a property of the q5 block rather than of any one decomposition.
	
	\FloatBarrier
	\begin{table}[H]
		\centering
		\singlespacing
		\begin{threeparttable}
			\caption{q5 alphas across daily size deciles}
			\label{tab:q5_daily_size_decile_alphas}
			\footnotesize
			\setlength{\tabcolsep}{4pt}
			\begin{tabular}{lrrrrrr}
				\toprule
				& \multicolumn{3}{c}{French deciles} & \multicolumn{3}{c}{OSAP deciles} \\
				\cmidrule(lr){2-4} \cmidrule(lr){5-7}
				Decile & $\alpha$ & $t(\alpha)$ & $p(\alpha)$ & $\alpha$ & $t(\alpha)$ & $p(\alpha)$ \\
				& (bp/yr) & & & (bp/yr) & & \\
				\midrule
				Smallest & 231.0 & 2.26 & 0.024 & 481.8 & 1.98 & 0.048 \\
				D02      &   0.9 & 0.01 & 0.988 & 410.1 & 2.09 & 0.037 \\
				D03      & 109.8 & 2.28 & 0.023 & 413.9 & 2.75 & 0.006 \\
				D04      &  81.4 & 1.75 & 0.080 & 164.9 & 1.53 & 0.125 \\
				D05      & 174.5 & 3.41 & $6.58 \times 10^{-4}$ & -18.9 & -0.25 & 0.799 \\
				D06      & 179.6 & 3.14 & 0.002 &  40.3 & 0.76 & 0.445 \\
				D07      & 251.3 & 4.48 & $7.43 \times 10^{-6}$ &  24.1 & 0.57 & 0.570 \\
				D08      & 230.1 & 4.37 & $1.27 \times 10^{-5}$ &  89.7 & 2.00 & 0.045 \\
				D09      & 118.1 & 2.38 & 0.018 & 216.9 & 4.39 & $1.12 \times 10^{-5}$ \\
				Largest  & -70.0 & -2.98 & 0.003 & -15.7 & -1.16 & 0.244 \\
				\bottomrule
			\end{tabular}
			\begin{tablenotes}[flushleft]
				\footnotesize
				\item Notes: The table reports individual q5 alpha estimates for daily value-weighted size deciles. Alphas are annualized basis points. HAC standard errors use 21 daily lags. Deciles are ordered from the smallest to the largest size portfolio within each source. The OSAP long--short portfolio is excluded.
			\end{tablenotes}
		\end{threeparttable}
	\end{table}
	\FloatBarrier
	
	Table~\ref{tab:q5_daily_size_decile_alphas} gives the q5-only cross-section, and it rules out a microcap or nonsynchronous-trading reading. If the pattern were such an artifact, the sharpest alphas should sit at the microcap end; instead they sit in the interior of the size distribution in both libraries---D07 and D08 in French ($t=4.48$, $4.37$), D03 and D09 in OSAP ($t=2.75$, $4.39$)---while the smallest decile is significant in French but only marginal in OSAP ($t=1.98$). The largest decile carries a negative q5 alpha in both libraries ($-70.0$ and $-15.7$ bp), the same sign q5 assigns to the body leg internally. The internal body--tail design agrees: were the pattern a small-stock artifact, it should be sharpest where the tail is most microcap-concentrated, but Table~\ref{tab:q5_body_tail_by_ratio} shows the q5 tail-alpha $t$-statistic peaks at 4.37 to 5.14 for split ratios 0.50 to 0.80, where the tail holds 50\% to 20\% of market value. So the external deciles and the internal decomposition locate the error in the same place---negative large-cap alpha, sharpest positive alphas in the interior---and detailed profiles differ across constructions but that coarse shape does not. Microstructure may move individual estimates but not that location; with the HAC-lag stability of Table~\ref{tab:q5_hac_sensitivity}, a pure microstructure reading is hard to sustain.
	
	\subsection{Factor-block ablation across model families}
	
	Table~\ref{tab:ablation_across_families} summarizes selected ablations---a within-family diagnostic, not a sample-controlled horse race---showing how the body--tail pattern changes when specific factor blocks are removed or restricted.
	
	\FloatBarrier
	\begin{table}[H]
		\centering
		\singlespacing
		\resizebox{\textwidth}{!}{
			\begin{threeparttable}
				\caption{Selected factor-block ablations across model families}
				\label{tab:ablation_across_families}
				\scriptsize
				\setlength{\tabcolsep}{3pt}
				\begin{tabular}{lllrrrrr}
					\toprule
					Family & Variant & Kept factors & Sample & Joint & Median $p_J$ & $\bar\alpha_A$ & $\bar\alpha_B$ \\
					&         &              &        & reject &              & (bp/yr) & (bp/yr) \\
					\midrule
					FF3 & Full FF3 & MKT+SMB+HML & 1926--2024 & 1/9 & 0.230 & 12.6 & -7.6 \\
					& MKT only & MKT & 1926--2024 & 2/9 & 0.103 & -22.0 & 164.1 \\
					& Drop SMB & MKT+HML & 1926--2024 & 1/9 & 0.328 & -8.5 & 101.3 \\
					& Drop HML & MKT+SMB & 1926--2024 & 1/9 & 0.273 & -2.9 & 65.4 \\
					\hline
					Carhart & Full Carhart & MKT+SMB+HML+UMD & 1926--2024 & 1/9 & 0.384 & 4.3 & 40.8 \\
					& MKT only & MKT & 1926--2024 & 2/9 & 0.092 & -23.3 & 168.6 \\
					& Drop HML & MKT+SMB+UMD & 1926--2024 & 8/9 & 0.019 & -15.0 & 130.7 \\
					& Drop UMD & MKT+SMB+HML & 1926--2024 & 1/9 & 0.257 & 11.9 & -6.4 \\
					\hline
					FF5 & Full FF5 & MKT+SMB+HML+RMW+CMA & 1963--2024 & 0/9 & 0.745 & -2.3 & -14.4 \\
					& MKT only & MKT & 1963--2024 & 0/9 & 0.667 & -21.2 & 98.4 \\
					& Drop SMB & MKT+HML+RMW+CMA & 1963--2024 & 2/9 & 0.086 & -48.3 & 222.0 \\
					& Drop RMW & MKT+SMB+HML+CMA & 1963--2024 & 3/9 & 0.450 & 10.4 & -61.6 \\
					& Drop CMA & MKT+SMB+HML+RMW & 1963--2024 & 0/9 & 0.853 & 0.7 & -26.9 \\
					\hline
					FF6 & Full FF6 & MKT+SMB+HML+RMW+CMA+UMD & 1963--2024 & 0/9 & 0.567 & -10.3 & 36.7 \\
					& MKT only & MKT & 1963--2024 & 0/9 & 0.667 & -21.2 & 98.4 \\
					& Drop SMB & MKT+HML+RMW+CMA+UMD & 1963--2024 & 7/9 & 0.030 & -55.8 & 270.6 \\
					& Drop UMD & MKT+SMB+HML+RMW+CMA & 1963--2024 & 0/9 & 0.745 & -2.3 & -14.4 \\
					& Drop RMW & MKT+SMB+HML+CMA+UMD & 1963--2024 & 0/9 & 0.790 & 1.1 & -4.0 \\
					\hline
					q5 & Full q5 & MKT+ME+IA+ROE+EG & 1967--2024 & 9/9 & $<0.001$ & -56.2 & 181.0 \\
					& MKT only & MKT & 1967--2024 & 0/9 & 0.768 & -18.9 & 81.5 \\
					& Drop EG & MKT+ME+IA+ROE & 1967--2024 & 2/9 & 0.098 & -26.8 & 106.6 \\
					& Drop ROE & MKT+ME+IA+EG & 1967--2024 & 6/9 & 0.001 & -51.0 & 151.8 \\
					& Drop ME & MKT+IA+ROE+EG & 1967--2024 & 9/9 & $<0.001$ & -143.7 & 629.1 \\
					\bottomrule
				\end{tabular}
				\begin{tablenotes}[flushleft]
					\footnotesize
					\item Notes: ``Joint reject'' counts 5\% Wald rejections across the nine body--tail split ratios. HAC standard errors use 21 daily lags.
				\end{tablenotes}
			\end{threeparttable}
		}
	\end{table}
	\FloatBarrier
	
	The traditional Fama--French specifications are broadly stable. Removing SMB from FF5 or FF6 raises body--tail rejections, so the size block absorbs part of the size-ranked decomposition rather than generating it: size enters on the stabilizing side. q5 behaves oppositely. Its market-only variant rejects at no cutoff while the full specification rejects at all nine, so adding the q5 nonmarket block turns a passing decomposition into a rejected one---the same direction seen in the external size deciles.
	
	Within the q5 block, the rejections track the profitability--growth side rather than size. Dropping EG collapses the pattern from 9/9 to 2/9 (median $p_J=0.098$) and dropping ROE leaves more in place at 6/9, so EG carries most of the effect and ROE the remainder; dropping ME, by contrast, not only keeps all nine rejections but amplifies the leg alphas (to $-143.7$ and $629.1$ bp), confirming once more that the source is not a missing size factor. The next subsection decomposes this block.
	
	\subsection{The ROE--EG block in q5}
	
	Table~\ref{tab:q5_ablation} examines q5 variants more closely. With the native market factor (Panel A), MKT-only, MKT+ME, and MKT+ME+IA produce no joint rejections; rejections appear only once ROE or EG enters, and the attenuation is largest when EG is removed---dropping EG cuts rejections from nine to two while dropping ROE leaves six---so EG carries most of the body--tail sensitivity, ROE second. This is the point of contact with Section~\ref{sec:span}, where EG was the single largest contributor to q5's mean--variance advantage over FF6, with an individual $\Delta\operatorname{SR}^2$ of 2.425: the factor that most expands the q5 opportunity set is the one most associated with the body--tail pricing error. The sensitivity is also specific to the q-family profitability construction, not to profitability factors in general---with expected growth absent, MKT+ME+ROE rejects at three ratios while MKT+ME+RMW rejects at none, and replacing ROE with the Fama--French RMW does not remove the pattern when EG remains. Panel B repeats the diagnostic with cMKT-based variants and reaches the same ranking, so it does not depend on the market factor. I read these comparisons as evidence about which block carries the pattern in this design, not that any factor is economically invalid.
	
	\FloatBarrier
	\begin{table}[H]
		\centering
		\singlespacing
			\begin{threeparttable}
				\caption{q5 factor-block diagnostics}
				\label{tab:q5_ablation}
				\footnotesize
				\begin{tabular}{llrrrrr}
					\toprule
					Variant & Kept factors & Joint & Median $p_J$ & $\bar\alpha_A$ & $\bar\alpha_B$ & Sum \\
					&              & reject &              & (bp/yr) & (bp/yr) & reject \\
					\midrule
					\multicolumn{7}{l}{\textit{Panel A: Native q5 market factor}} \\
					Full q5 & MKT+ME+IA+ROE+EG & 9/9 & $<0.001$ & -56.2 & 181.0 & 8/9 \\
					MKT only & MKT & 0/9 & 0.768 & -18.9 & 81.5 & 0/9 \\
					MKT+ME & MKT+ME & 0/9 & 0.530 & 12.9 & -77.3 & 4/9 \\
					MKT+ME+IA & MKT+ME+IA & 0/9 & 0.429 & 11.2 & -72.1 & 3/9 \\
					MKT+ME+ROE & MKT+ME+ROE & 3/9 & 0.094 & -27.0 & 110.1 & 4/9 \\
					MKT+ME+EG & MKT+ME+EG & 6/9 & 0.008 & -48.8 & 144.7 & 6/9 \\
					MKT+ME+RMW & MKT+ME+RMW & 0/9 & 0.809 & -3.7 & -15.7 & 0/9 \\
					Drop EG & MKT+ME+IA+ROE & 2/9 & 0.098 & -26.8 & 106.6 & 5/9 \\
					Drop ROE & MKT+ME+IA+EG & 6/9 & 0.001 & -51.0 & 151.8 & 6/9 \\
					ROE to RMW & MKT+ME+IA+RMW+EG & 8/9 & $<0.001$ & -53.5 & 161.4 & 7/9 \\
					\hline
					\multicolumn{7}{l}{\textit{Panel B: Production market return (cMKT)}} \\
					Full cMKT q5 & cMKT+ME+IA+ROE+EG & 9/9 & $<0.001$ & -50.7 & 187.7 & 9/9 \\
					cMKT+ME+IA & cMKT+ME+IA & 1/9 & 0.247 & 11.6 & -70.6 & 2/9 \\
					cMKT+ME+EG & cMKT+ME+EG & 7/9 & 0.002 & -44.2 & 150.2 & 7/9 \\
					cMKT+ME+RMW & cMKT+ME+RMW & 1/9 & 0.591 & -2.2 & -13.3 & 0/9 \\
					ROE to RMW & cMKT+ME+IA+RMW+EG & 8/9 & $<0.001$ & -47.9 & 168.3 & 8/9 \\
					\bottomrule
				\end{tabular}
				\begin{tablenotes}[flushleft]
					\footnotesize
					\item Notes: q5 ablations over 1967--2024. ``Joint reject'' and ``Sum reject'' count 5\% rejections across the nine body--tail split ratios; alphas are annualized basis points; HAC standard errors use 21 daily lags. Panel A uses the native q5 market factor. Panel B replaces it with the production market return cMKT, for which net-alpha $p$-values are omitted because the net market is mechanically spanned by cMKT.
				\end{tablenotes}
			\end{threeparttable}
	\end{table}
	\FloatBarrier
	
	The ablations therefore place the body--tail sensitivity in the q5 profitability--growth block. ME stabilizes the decomposition---dropping it strengthens the rejection and sharply raises the tail alpha---while the pattern is most pronounced when ROE and EG are present, and the cMKT panel gives the same ranking, ruling out the market factor as the driver. The interpretation is deliberately narrow: the ablations identify the block associated with the pricing-error pattern in this design, not that the factors are misspecified in a mean--variance sense, where Section~\ref{sec:span} showed the same block is q5's main strength. That a single block is the source of both the spanning advantage and the body--tail pricing error is precisely the separation between the two criteria that motivates the paper.
	
	\subsection{Loading--premium decomposition}
	
	The ablations show where to look; I next decompose the tail-minus-body spread. For each split, let $\widehat{c}_k=(\widehat{\beta}_{T,k}-\widehat{\beta}_{B,k})\,\overline{f}_k$, so the average tail-minus-body spread splits into an alpha spread and a factor-implied spread,
	\[
	\overline{R}_T-\overline{R}_B
	=\widehat{\alpha}_{T-B}+\sum_k \widehat{c}_k ,
	\]
	with all nine pairs reconstructing the production market numerically and all identity checks passing.
	
	\FloatBarrier
	\begin{table}[H]
		\centering
		\singlespacing
		\resizebox{\textwidth}{!}{
			\begin{threeparttable}
				\caption{q5 loading--premium decomposition of tail-minus-body spreads}
				\label{tab:q5_loading_premium_decomposition}
				\footnotesize
				\begin{tabular}{llrrrrrrrr}
					\toprule
					Model & Window & Actual & Factor & Alpha & MKT & ME & IA & ROE & EG \\
					& & (bp/yr) & (bp/yr) & (bp/yr) & (bp/yr) & (bp/yr) & (bp/yr) & (bp/yr) & (bp/yr) \\
					\midrule
					q5 & Full period & 88.3 & -148.9 & 237.2 & -28.0 & 170.9 & 0.9 & -166.5 & -126.2 \\
					cMKT--q5 & Full period & 88.3 & -150.1 & 238.4 & -28.6 & 170.9 & 0.7 & -166.4 & -126.6 \\
					q5 & Rolling 5y & 76.6 & -148.8 & 225.4 & -32.2 & 147.1 & 15.7 & -139.1 & -140.2 \\
					cMKT--q5 & Rolling 5y & 76.6 & -149.4 & 226.0 & -32.5 & 147.0 & 15.3 & -139.0 & -140.1 \\
					\bottomrule
				\end{tabular}
				\begin{tablenotes}[flushleft]
					\footnotesize
					\item Notes: The table averages across the nine split ratios. ``Actual'' is the realized tail-minus-body spread. ``Factor'' is the sum of factor-loading contributions. ``Alpha'' is Actual minus Factor. Rolling 5y rows average over all rolling five-year windows and split ratios.
				\end{tablenotes}
			\end{threeparttable}
		}
	\end{table}
	\FloatBarrier
	
	Table~\ref{tab:q5_loading_premium_decomposition} reports the decomposition, and the sign pattern is its main result. The realized tail-minus-body spread is positive (88.3 bp) while the q5 factor-implied spread is negative ($-148.9$ bp), leaving a large positive residual alpha spread (237.2 bp). The components show why: ME contributes positively (170.9 bp), consistent with the small-stock exposure of the tail, but ROE and EG contribute negatively ($-166.5$ and $-126.2$ bp) and more than offset it. The factor-implied spread is negative not despite the size block but against it---the profitability--growth block pulls the implied tail return below the body, while the realized tail return is above it.
	
	\FloatBarrier
	\begin{table}[H]
		\centering
		\singlespacing
		\begin{threeparttable}
			\caption{Sign persistence in rolling five-year windows}
			\label{tab:q5_rolling5_sign_persistence}
			\footnotesize
			\setlength{\tabcolsep}{3pt}
			\begin{tabular}{lrrrrr}
				\toprule
				Model & Observations & $\Pr(\alpha_{T-B}>0)$ (\%) & $\Pr(c_{ME}>0)$ (\%) & $\Pr(c_{ROE}<0)$ (\%) & $\Pr(c_{EG}<0)$ (\%) \\
				\midrule
				q5 & 5733 & 84.8 & 58.7 & 96.2 & 90.7 \\
				cMKT--q5 & 5733 & 84.8 & 58.7 & 96.2 & 90.7 \\
				\bottomrule
			\end{tabular}
			\begin{tablenotes}[flushleft]
				\footnotesize
				\item Notes: Observations are rolling five-year window by split-ratio pairs. There are 637 windows and nine split ratios for each model.
			\end{tablenotes}
		\end{threeparttable}
	\end{table}
	\FloatBarrier
	
	The structure persists beyond the full sample. In rolling five-year windows (Table~\ref{tab:q5_rolling5_sign_persistence}) the alpha spread is positive in about 85\% of split-window observations and the ROE and EG contributions are negative in most (96.2\% and 90.7\%); nonoverlapping phases (Table~\ref{tab:q5_nonoverlap_phase_regressions}) give the same ordering without the overlapping-window concern, with larger negative EG and ROE contributions associated with larger alpha spreads in 98.3\% and 96.7\% of phases; and rolling-window HAC regressions (Table~\ref{tab:q5_contribution_hac_regressions}) confirm that alpha spreads widen as the ROE and EG contributions turn more negative, while ME enters with the opposite, stabilizing sign ($R^2=0.543$). The full-sample decomposition is thus not the product of one period or one split ratio, and it fixes the time-variation result of Section~\ref{sec:subperiod}: the pattern is active exactly when the negative ROE and EG contributions are large, the same channel through which the rejection switches on and off across decades.
	
	\FloatBarrier
	\begin{table}[H]
		\centering
		\singlespacing
		\begin{threeparttable}
			\caption{Nonoverlapping phase regressions for alpha spreads}
			\label{tab:q5_nonoverlap_phase_regressions}
			\footnotesize
			\setlength{\tabcolsep}{3pt}
			\begin{tabular}{llrrrrr}
				\toprule
				Model & Regressor & Phases & Positive slope (\%) & Median $R^2$ & Median corr. & Significant (\%) \\
				\midrule
				q5 & $c_{ME}$ & 60 & 86.7 & 0.147 & 0.384 & 78.3 \\
				q5 & $c_{IA}$ & 60 & 53.3 & 0.043 & 0.032 & 58.3 \\
				q5 & $-c_{ROE}$ & 60 & 96.7 & 0.048 & 0.218 & 55.0 \\
				q5 & $-c_{EG}$ & 60 & 98.3 & 0.095 & 0.309 & 78.3 \\
				cMKT--q5 & $c_{ME}$ & 60 & 86.7 & 0.147 & 0.383 & 76.7 \\
				cMKT--q5 & $c_{IA}$ & 60 & 53.3 & 0.043 & 0.030 & 56.7 \\
				cMKT--q5 & $-c_{ROE}$ & 60 & 96.7 & 0.047 & 0.218 & 55.0 \\
				cMKT--q5 & $-c_{EG}$ & 60 & 98.3 & 0.093 & 0.306 & 76.7 \\
				\bottomrule
			\end{tabular}
			\begin{tablenotes}[flushleft]
				\footnotesize
				\item Notes: Each phase uses nonoverlapping five-year windows. The dependent variable is the alpha spread across split ratios within a phase. ``Significant'' is the share of phases with a 5\% significant slope. Regressors with a minus sign are multiplied by $-1$ so that a positive slope means a larger negative contribution is associated with a larger alpha spread.
			\end{tablenotes}
		\end{threeparttable}
	\end{table}
	\FloatBarrier
	
	Nonoverlapping phases give the same message without the overlapping-window concern: larger negative EG and ROE contributions are associated with larger alpha spreads in 98.3\% and 96.7\% of phases.
	
	\FloatBarrier
	\begin{table}[H]
		\centering
		\singlespacing
		\begin{threeparttable}
			\caption{Rolling-window HAC regressions of alpha spreads on factor contributions}
			\label{tab:q5_contribution_hac_regressions}
			\footnotesize
			\begin{tabular}{lrrrrr}
				\toprule
				Model & $c_{ME}$ & $c_{IA}$ & $c_{ROE}$ & $c_{EG}$ & $R^2$ \\
				\midrule
				q5 & 0.253*** (4.20) & -0.475 (-1.10) & -0.945*** (-6.06) & -0.292** (-2.46) & 0.543 \\
				cMKT--q5 & 0.253*** (4.21) & -0.462 (-1.07) & -0.949*** (-6.05) & -0.283** (-2.38) & 0.543 \\
				\bottomrule
			\end{tabular}
			\begin{tablenotes}[flushleft]
				\footnotesize
				\item Notes: The dependent variable is the rolling five-year alpha spread averaged across split ratios. Coefficients are reported with Newey--West $t$-statistics in parentheses. The HAC lag is 60 months. $^{*}$, $^{**}$, and $^{***}$ denote 10\%, 5\%, and 1\% significance.
			\end{tablenotes}
		\end{threeparttable}
	\end{table}
	\FloatBarrier
	
	The mechanism behind the headline pattern follows directly. The evidence does not support a missing-size-factor reading: in this design the size block contributes on the stabilizing side, and the ROE--EG block is the source of the negative q5 factor-implied tail-minus-body spread that the positive realized spread must overcome. q5 assigns the size-ranked tail too low a factor-implied return---because the tail loads on ROE and EG in a way that, at the q5 premia, implies a lower return than the body---so a positive realized tail-minus-body spread is left as a positive alpha spread. The conclusion is diagnostic and design-specific: it describes how q5 allocates factor-implied returns across this size-ranked decomposition, not a general failure of the profitability or expected-growth factors, which on the spanning criterion are q5's strongest components.
	
	\section{Discussion}
	\label{sec:discuss}
	
	\subsection{What the cap-axis diagnostic adds}
	
	The aggregate market regression is a useful sanity check but a weak endpoint when the candidate model already contains a market factor: a value-weighted market can price at zero while its components carry offsetting errors. The body--tail diagnostic keeps that aggregate relation intact and asks instead whether a model that prices the compressed market also prices an economically ordered decomposition of it.
	
	The decomposition axis is not arbitrary. Cumulative market capitalization is both the rule that weights each stock in the aggregate market and the rule that orders the legs, so the cap-axis cut aligns the weighting variable with the sorting variable---the value-weighted body and tail reconstruct the production market return exactly, and the model is asked to price those components while still pricing their recombination. This is what makes the test internally anchored rather than an external anomaly library bolted onto the market: the market portfolio is at once the object reconstructed and the source of the diagnostic split. A model that contains a market factor can pass the aggregate regression almost mechanically, especially when the test return is close to the provider's own market factor; the cap-axis split makes that pass less final by treating the market as an object with internal anatomy. If the aggregate alpha is small because errors offset across capitalization-ranked components, the split reveals the offset while preserving the identity.
	
	Because the recombination identity holds for every model alike, it cannot itself explain why one model leaves a leg pattern and others do not---the diagnostic content is the cross-model contrast. The matched random split sharpens it: when the same market is divided at comparable target ratios but the size-ranked rule is removed, the daily q5 rejection rate falls close to nominal. The daily q5 result is therefore not a generic split effect but a property of the size-ranked decomposition, invisible to aggregate alpha tests that compress the cross-section into one return.
	
	\subsection{Relation to spanning}
	
	The diagnostic is not a substitute for the spanning criterion of \citet{BS17} and \citet{BS18}, and the two answer different questions: spanning asks whether one traded-factor model expands the mean--variance opportunity set relative to another, while the body--tail test asks whether a model prices an economically ordered decomposition of a market it already prices in aggregate. The distinction has force precisely because q5 is strong on spanning---its profitability and expected-growth factors expand the frontier well beyond the Fama--French block---yet leaves a systematic daily leg pattern. A model can improve maximum-Sharpe performance and still misprice a particular test-asset geometry.
	
	There is no contradiction. For an ideal SDF, pricing all assets and spanning the tangency portfolio coincide; for a low-dimensional approximation they separate, so a factor block can capture a high-Sharpe direction and, through the same loadings and premia, generate pricing errors on a particular decomposition. The separation is sharpest in the q5 profitability--growth block: expected growth is a major contributor to q5's spanning advantage (Section~\ref{sec:span}), yet removing it attenuates the daily body--tail imbalance most (Section~\ref{sec:diagnostics}). The block that improves one evaluation object worsens another. This is why the paper's claim is not that q5 is globally inferior---it remains the strongest candidate on spanning, and it becomes much less fragile under monthly aggregation---but the narrower, more diagnostic one that aggregate fit, spanning, and internal-decomposition pricing are distinct empirical objects on which the same model can stand differently.
	
	\subsection{Frequency dependence}
	
	The monthly results are central to that claim, because they show the daily ranking is not frequency-invariant. At the daily frequency q5 is the clear outlier; after compounding the same CRSP legs to calendar months and using monthly factor returns, its sharp joint rejection largely attenuates while the sign pattern and alpha-sum imbalance remain, and the most visible joint-test weakness shifts toward FF3 in the common sample. As Section~\ref{sec:monthly} argues, this is not mere power loss: a uniform sample-size effect would shrink every q5 statistic in step, whereas here the joint test collapses while the alpha-sum test does not, so aggregation changes the \emph{shape} of the signal, not only its scale.
	
	The reframing is the contribution. The object is not a frequency-free model failure but frequency-dependent internal consistency: a model can be fragile under a daily implementation and much less so under a monthly one, and another model can move the opposite way. This is economically plausible. The CRSP panel is observed daily, so the daily implementation uses the native frequency of the return data and the published daily factor series; but accounting-based characteristics, slower rebalancing, and lower-frequency premia may cohere more naturally with monthly returns, while daily estimation can expose short-horizon timing and covariance mismatches that monthly averaging removes. The monthly evidence therefore does not undercut the daily evidence---the daily q5 result is valid for a daily CRSP implementation of the published daily series---but it warns against reading the daily ranking as a frequency-free property of the model.
	
	The same logic disciplines the mechanism. The daily diagnostics locate the q5 pattern on the nonmarket side---the market factor and $cMKT$ substitution leave it intact, and within the block the profitability--growth side carries it, with the loading--premium decomposition showing that ROE and EG drive the q5 factor-implied tail-minus-body spread negative against a positive realized spread (Section~\ref{sec:diagnostics}). The size block stabilizes rather than generates the pattern, so it is not a missing-size-factor story. This identification is daily-specific: the same block's strength and ranking shift under monthly aggregation. Consistently, the daily pattern is not constant over time---strongest from the 1980s through the mid-2000s, weak recently, and absent entirely in some regimes---a profile that switches on and off as the ROE/EG premia wax and wane, which a fixed mechanical split artifact would not do but a factor-block mechanism can. The reading is descriptive, not causal: body--tail alpha depends jointly on factor means, loadings, covariances, and sample composition, and a weaker expected-growth premium is consistent with weaker recent rejections without being the sole driver.
	
	\subsection{Scope and interpretation}
	
	The evidence is specific to an investible CRSP common-stock portfolio, the universe filters used here, annual June formation, dynamic value-weighted recombination, and the daily and monthly factor implementations tested. The robustness checks reduce concern about particular implementation choices but do not imply invariance across every market definition, rebalancing rule, or asset universe. The external size deciles reinforce this: French and OSAP do not give identical rankings, which is not a weakness but part of the message---test-asset construction matters, and the body--tail design is useful because it makes one construction transparent and anchors it to the market portfolio itself. Other economically ordered decompositions are possible and would be useful extensions; the present design is deliberately narrow, studying one transparent cut whose sorting variable is also the market's weighting variable.
	
	The main conclusion is therefore not that one model is universally best or worst, but that factor models are conditional approximations whose empirical standing depends on the evaluation object: aggregate market alpha, factor spanning, body--tail pricing, random-split placebo behavior, external size portfolios, and observation frequency. The body--tail diagnostic adds one controlled way to see this conditionality---showing that a model can price the aggregate market, expand the mean--variance frontier, and still leave frequency-dependent pricing errors inside the market's own size-ranked anatomy.
	
	\section{Conclusion}
	\label{sec:conclusion}
	
	I study whether a factor model that prices the aggregate market also prices market-derived portfolios formed from the same investible universe. Because the aggregate regression is a limited endpoint when the model already contains a market factor, I decompose an investible CRSP market portfolio into body and tail legs along the cumulative-market-capitalization axis---tradable, value-weighted portfolios whose dynamic recombination reconstructs the aggregate market return. The diagnostic holds the aggregate relation fixed and asks whether pricing errors remain hidden inside the market portfolio itself.
	
	The daily evidence shows a sharp separation. Every model prices the aggregate portfolio with a small alpha and very high explanatory power, yet once the same market is split along the cap axis, q5 rejects the joint zero-alpha restriction at all nine ratios with negative body and positive tail alphas, while CAPM, Carhart, FF5, and FF6 are far more stable and FF3 is intermediate. The recombination identity holds for every model, so the pattern cannot be a mechanical consequence of splitting, and the matched random split---which removes only the size-ranked assignment rule---collapses the q5 rejection rate to nominal. The daily result is thus a cap-axis pricing pattern hidden by the aggregate regression, and the daily diagnostics trace it to q5's nonmarket profitability--growth block rather than to its market factor or to microstructure.
	
	The monthly evidence changes the ranking and is central to the interpretation. Compounding the same legs to calendar months substantially weakens the q5 joint rejection---the pattern attenuates rather than disappearing, surviving mainly through the alpha-sum restriction---while the most visible joint-test weakness shifts toward FF3 in the common sample. Because the q5 joint test collapses while its alpha-sum test does not, this is a change in the shape of the signal, not uniform power loss. The diagnostic is therefore not only model-dependent but frequency-dependent: factor models are frequency-conditional approximations of the SDF, and internal pricing diagnostics can depend on the frequency at which test-asset and factor returns are evaluated.
	
	None of this overturns the spanning criterion of \citet{BS17} and \citet{BS18}. q5 remains the strongest candidate on spanning, and the same profitability--growth block that drives that advantage is the one associated with the daily cap-axis imbalance---the separation an approximate SDF permits between pricing all assets and spanning the tangency portfolio. The broader contribution is methodological: by aligning the market's weighting rule with the decomposition axis, the cap-axis diagnostic uses the market portfolio itself as a testing device, confronting the model not with an external anomaly library but with tradable components of the market it already prices. Passing the aggregate regression therefore does not guarantee pricing consistency across market-derived test assets. The conclusion is not that any one model is universally dominated, but that a market factor can hide offsetting internal pricing errors whose location and strength depend on both the economic axis used to decompose the market and the frequency at which the model is evaluated.
	
	\paragraph{Funding}
	This research did not receive any specific grant from funding agencies in the public, commercial, or not-for-profit sectors.
	
	\paragraph{Declaration of AI usage} 
	During the preparation of this manuscript, the author used ChatGPT (OpenAI) and Claude (Anthropic) for language refinement and structural clarity. 	All outputs were reviewed and edited by the author, who takes full responsibility for the content.
	
	\paragraph{Declaration of interest}
	The author declares no competing interests.
	


\newpage
	\begin{thebibliography}{99}
		
		\singlespacing
			
		\bibitem[Gibbons et al.(1989)]{GRS89}
		Gibbons, M. R., Ross, S. A., \& Shanken, J. (1989).
		A test of the efficiency of a given portfolio.
		\textit{Econometrica}, 57(5), 1121--1152.
		\url{https://www.jstor.org/stable/1913625}
		
		\bibitem[Lo and MacKinlay(1990)]{LM90}
		Lo, A. W., \& MacKinlay, A. C. (1990).
		Data-snooping biases in tests of financial asset pricing models.
		\textit{The Review of Financial Studies}, 3(3), 431--467.
		\url{https://doi.org/10.1093/rfs/3.3.431}
		
		\bibitem[Hansen and Jagannathan(1997)]{HJ97}
		Hansen, L. P., \& Jagannathan, R. (1997).
		Assessing specification errors in stochastic discount factor models.
		\textit{The Journal of Finance}, 52(2), 557--590.
		\url{https://doi.org/10.1111/j.1540-6261.1997.tb04813.x}
		
		\bibitem[Cochrane(2005)]{Cochrane05}
		Cochrane, J. H. (2005).
		\textit{Asset Pricing} (Revised ed.).
		Princeton University Press.
		\url{https://www.johnhcochrane.com/asset-pricing}
		
		\bibitem[Lewellen et al.(2010)]{LNS10}
		Lewellen, J., Nagel, S., \& Shanken, J. (2010).
		A skeptical appraisal of asset-pricing tests.
		\textit{Journal of Financial Economics}, 96(2), 175--194.
		\url{https://doi.org/10.1016/j.jfineco.2009.09.001}
		
		\bibitem[Barillas and Shanken(2017)]{BS17}
		Barillas, F., \& Shanken, J. (2017).
		Which alpha?
		\textit{The Review of Financial Studies}, 30(4), 1316--1338.
		\url{https://doi.org/10.1093/rfs/hhw101}
		
		\bibitem[Barillas and Shanken(2018)]{BS18}
		Barillas, F., \& Shanken, J. (2018).
		Comparing asset pricing models.
		\textit{The Journal of Finance}, 73(2), 715--754.
		\url{https://doi.org/10.1111/jofi.12607}
		
		\bibitem[Kozak et al.(2018)]{KNS18}
		Kozak, S., Nagel, S., \& Santosh, S. (2018).
		Interpreting factor models.
		\textit{The Journal of Finance}, 73(3), 1183--1223.
		\url{https://doi.org/10.1111/jofi.12612}
		
		\bibitem[Chen and Zimmermann(2022)]{CZ22}
		Chen, A. Y., \& Zimmermann, T. (2022).
		Open source cross-sectional asset pricing.
		\textit{Critical Finance Review}, 11(2), 207--264.
		\url{https://doi.org/10.1561/104.00000112}
		
		\bibitem[Giglio et al.(2025)]{GXZ25}
		Giglio, S., Xiu, D., \& Zhang, D. (2025).
		Test assets and weak factors.
		\textit{The Journal of Finance}, 80(1), 259--319.
		\url{https://doi.org/10.1111/jofi.13415}
		
		\bibitem[Shin(2026)]{Shin26}
		Shin, U. (2026).
		Which portfolios? The construction dependence of factor model performance.
		\textit{arXiv preprint arXiv:2606.19550}.
		\url{https://doi.org/10.48550/arXiv.2606.19550}
		
		
		\bibitem[Jegadeesh and Titman(1993)]{JT93}
		Jegadeesh, N., \& Titman, S. (1993).
		Returns to buying winners and selling losers: Implications for stock market efficiency.
		\textit{The Journal of Finance}, 48(1), 65--91.
		\url{https://doi.org/10.1111/j.1540-6261.1993.tb04702.x}
		
		\bibitem[Carhart(1997)]{Carhart97}
		Carhart, M. M. (1997).
		On persistence in mutual fund performance.
		\textit{The Journal of Finance}, 52(1), 57--82.
		\url{https://doi.org/10.1111/j.1540-6261.1997.tb03808.x}
		
		\bibitem[Fama and French(1992)]{FF92}
		Fama, E. F., \& French, K. R. (1992).
		The cross-section of expected stock returns.
		\textit{The Journal of Finance}, 47(2), 427--465.
		\url{https://doi.org/10.1111/j.1540-6261.1992.tb04398.x}
		
		\bibitem[Fama and French(1993)]{FF93}
		Fama, E. F., \& French, K. R. (1993).
		Common risk factors in the returns on stocks and bonds.
		\textit{Journal of Financial Economics}, 33(1), 3--56.
		\url{https://doi.org/10.1016/0304-405X(93)90023-5}
		
		\bibitem[Fama and French(2015)]{FF15}
		Fama, E. F., \& French, K. R. (2015).
		A five-factor asset pricing model.
		\textit{Journal of Financial Economics}, 116(1), 1--22.
		\url{https://doi.org/10.1016/j.jfineco.2014.10.010}
		
		\bibitem[Fama and French(2018)]{FF18}
		Fama, E. F., \& French, K. R. (2018).
		Choosing factors.
		\textit{Journal of Financial Economics}, 128(2), 234--252.
		\url{https://doi.org/10.1016/j.jfineco.2018.02.012}
		
		
		\bibitem[Hou et al.(2015)]{HXZ15}
		Hou, K., Xue, C., \& Zhang, L. (2015).
		Digesting anomalies: An investment approach.
		\textit{The Review of Financial Studies}, 28(3), 650--705.
		\url{https://doi.org/10.1093/rfs/hhu068}
		
		\bibitem[Hou et al.(2019)]{HMXZ19}
		Hou, K., Mo, H., Xue, C., \& Zhang, L. (2019).
		Which factors?
		\textit{Review of Finance}, 23(1), 1--35.
		\url{https://doi.org/10.1093/rof/rfy032}
		
		\bibitem[Hou et al.(2020)]{HXZ20}
		Hou, K., Xue, C., \& Zhang, L. (2020).
		Replicating anomalies.
		\textit{The Review of Financial Studies}, 33(5), 2019--2133.
		\url{https://doi.org/10.1093/rfs/hhy131}
		
		\bibitem[Hou et al.(2021)]{HMXZ21}
		Hou, K., Mo, H., Xue, C., \& Zhang, L. (2021).
		An augmented q-factor model with expected growth.
		\textit{Review of Finance}, 25(1), 1--41.
		\url{https://doi.org/10.1093/rof/rfaa004}
		
		\bibitem[Hou et al.(2024)]{HMXZ24}
		Hou, K., Mo, H., Xue, C., \& Zhang, L. (2024).
		The economics of security analysis.
		\textit{Management Science}, 70(1), 164--186.
		\url{https://doi.org/10.1287/mnsc.2022.4640}
		
		
		\bibitem[CRSP(2026)]{CRSP}
		Center for Research in Security Prices, LLC. (2026).
		\textit{CRSP US Stock Databases} [Data set].
		Accessed via Wharton Research Data Services, June 5, 2026.
		\url{https://www.crsp.org/research/}
		
		\bibitem[French(2026)]{FrenchDataLibrary}
		French, K. R. (2026).
		\textit{Kenneth R. French Data Library} [Data set].
		Accessed May 5, 2026.
		\url{https://mba.tuck.dartmouth.edu/pages/faculty/ken.french/data_library.html}
		
		\bibitem[Global-q.org(2026)]{globalqFactors}
		Global-q.org. (2026).
		\textit{Factors and testing portfolios} [Data set].
		Accessed May 5, 2026.
		\url{https://global-q.org/factors.html}
		
		\bibitem[Open Source Asset Pricing(2026)]{OSAP}
		Open Source Asset Pricing. (2026).
		\textit{Open Source Asset Pricing} [Data set].
		Accessed June 24, 2026.
		\url{https://www.openassetpricing.com}
		
	\end{thebibliography}
\end{document}